\documentclass[manuscript]{aastex}

\makeatletter
\newcommand*{\rom}[1]{\expandafter\@slowromancap\romannumeral #1@}
\makeatother
\slugcomment{Not to appear in Nonlearned J., 45.}

\shorttitle{Collapsed Cores in Globular Clusters}
\shortauthors{Djorgovski et al.}
\usepackage{subfigure}
\usepackage{rotating}
\usepackage{longtable}
\usepackage{graphics}
\usepackage{amsmath,bm}

\begin{document}
\title{An Systematic Analysis of Stellar Population in the Host Galaxies of SDSS Type I QSOs  }
\author{Jun-Jie Jin\altaffilmark{1,2}, Yi-Nan Zhu\altaffilmark{1}, Xian-Min Meng\altaffilmark{1}, Feng-Jie Lei\altaffilmark{1,2}, Hong Wu\altaffilmark{1}}
\altaffiltext{1}{Key Laboratory of Optical Astronomy, National Astronomical Observatories, Chinese
	Academy of Sciences, Beijing 100012, P.R. China; zyn@bao.ac.cn}
\altaffiltext{2}{School of Astronomy and Space Science University of Chinese Academy of Sciences,
	Beijing 100049, China; jjjin@bao.ac.cn}

\newcounter{RomanNumber}
\newcommand{\MyRoman}[1]{\setcounter{RomanNumber}{#1}\Roman{RomanNumber}}

\begin{abstract}
We investigate the relationship between host galaxies' stellar content and active galactic nuclei (AGN) for 
optically selected QSOs with z$<$0.5. There are total 82 QSOs we select from Sloan Digital Sky Survey (SDSS) . These 82 QSOs both have Wide-field Infrared Survey Explorer (WISE) data and measurable stellar content.
 With the help of the 
stellar population synthesis code STARLIGHT, we determine the luminosity fraction of AGN ,stellar population ages 
 and star-formation history (SFH) of host galaxies. We find out there is a correlation 
  between the star formation history and AGN property which suggests a possible delay from star formation to AGN. This probably indicates that the AGN activity correlate with the star formation activity which consistent with a co-evolution scheme for black hole and host galaxies.

\end{abstract}

\section{Introduction}
The connection between active galactic nuclei (AGN) and starburst (SB) activity  in galaxies have been proposed for a long time, going back to the first discovery of Ultraluminous Infrared Galaxies (ULIRGs) . Many effort has been made to drive out what is the dominant energy output mechanism \citep{1972ApJ...176L..95R,1979ARA&A..17..477R,1988ApJ...325...74S,1991MNRAS.252..593Z}. More and more works have provided evidence that ULIRGs are powerd by a mixture of SB and AGN  \citep{1972ApJ...178..623T,1977ARA&A..15..437T,1988ApJ...325...74S,1991ApJ...370L..65B,1992ARA&A..30..705B,1996ApJ...471..115B,1991MNRAS.252..593Z,1998A&AS..127..521W,1998A&AS..132..181W,1999A&A...349..735Z,2001AJ....122...63C,2002ApJ...564..196X,2005MNRAS.361..776S,2006ChJAA...6..197C}. \cite{1988ApJ...325...74S} has proposed a evolution scenario where two gas-rich spirals merge first and then drive gas into the merger center, triggering nuclear SB before the ignition of a dust-enshrouded AGN. When the dust has been consumed or swept away by the strong outflow from AGN and supernovae, an optical quasar would appear. Meanwhile these strong outflow also quench the star formation and the growth of black hole. Eventually, this evolution path provides a plausible explanation for the tight correlation between black hole mass and bugle of their host galaxies \citep{1998AJ....115.2285M,2000ApJ...539L..13G,2000ApJ...539L...9F,2002ApJ...574..740T,2001AIPC..586..363K,2001MNRAS.320L..30M}. Many studies also show that the black hole (BH) growth and star formation history has similar evolution \citep{2004ApJ...615..209H,2008ApJ...679..118S,2010MNRAS.401.2531A}. By the way, the works of \cite{1999ApJ...524...65T} and \cite{2008ApJ...683L.119B} shows that minor merges are also able to trigger the nuclear activity. Recent studies \citep{2007ApJ...671.1388D,2012MNRAS.420L...8H,2013ApJ...772..132C,2016ChA&A..40..291Z,2016MNRAS.462.2246B} also suggested that AGN and SB activity may not contemporaneous and there is a time gap between AGN activity and star burst, but it is still a controversial argument \citep{1998AJ....115.2285M,2000ApJ...539L..13G,2005Natur.433..604D}.

In a word, at least one of the consequence of major-merger evolutionary scenarios is the existence of object which have a phase with both luminous quasar activity and post-starburst (or on-going star formation) signatures. So it is important to investigate the stellar population in QSOs host galaxies. In the work of \cite{1999MNRAS.308..377M}, they presented the HST imaging study and found that the R-K colors of host galaxies is consistent with mature stellar population. \cite{2001MNRAS.323..308N} fitted deep off-nuclear optical spectra of QSOs and found that the host galaxies are dominated by old stars. On the other hand, the close relationship between gas-rich mergers and nuclear activity was supported by several studies, which show a existence of recent star bursts in the host galaxies of QSOs. \cite{2003MNRAS.346.1055K} examined the properties of the host galaxies of narrow-line AGN and found that the host of high-luminosity AGN has much younger mean stellar age.

 \cite{2017MNRAS.472.4382R} presented the characterization of the first 62 MaNGA (Mapping Nearby Galaxies at the Apache Point Observatory) AGN hosts and found that for more luminous AGNs the contribution of younger stellar populations to the optical emission is larger than for low-luminosity ones. \cite{2018RMxAA..54..217S} presented the properties of a sample of 98 AGN host galaxies (both type-II and type-I) and found that AGN hosts are in the transition stage between star-forming and non-star-forming galaxies. There are also many other works that found similar results \citep{2001ApJ...555..719C,2004ApJ...614..586S,2007MNRAS.378...83L,2007MNRAS.378...23J,2010MNRAS.408..713W}.

However, the overwhelming luminosity of QSOs compared with the host galaxies is a difficult challenge because both the AGN's continuum and broad line will serious weaken the stellar features. Many remarkable works have been made over the past few years (as mentioned above) for their own motivations. For example, many works limited their sample to obscured AGNs whose centers are obscured by large amounts of dust so that their host galaxies can be studied  \citep{2007ApJ...671.1388D,2003MNRAS.346.1055K,2005MNRAS.356..480T,2000AJ....120.1579Y,2004ApJ...613..109H}. Some works present an off-axis observation to avoid QSOs contamination \citep{2001MNRAS.323..308N,2013ApJ...772..132C}, but this method may miss the young stellar population that may be present in the center of the galaxies. By performing the deconvolution method on 2-D spectra \citep{1998ApJ...494..472M}, \cite{2007MNRAS.378...83L} separated the individual spectra of QSOs and their host galaxies. There are also many works about post star-burst QSOs (PSQs) \citep{2000AJ....119...59C,2013ApJ...762...90C,2013ApJ...772...28W}, which have prominent Balmer absorptions from A-type stars, but one must note that because of the definition, these objects were selected to have moderate-age stellar population by design. 

Because the relationship between QSOs and their host galaxies has an important consequences in our understanding of galaxies evolution, it is necessary to investigate the QSOs' host stellar population. In this paper we perform an extensive and statistical analyses for source selected from Sloan Digital Sky Survey (SDSS) \citep{2000AJ....120.1579Y,2002AJ....123..485S}. We effectively select the objects with distinct stellar feature 
for the first time and study their stellar population, mid-IR color as well as AGN properties. We describe the sample selection and data reductions in Section 2. The method for decomposing AGNs and stellar population along with star formation history (SFH) in Section 3. The outputted results and the properties of these QSOs as well as stellar population of host galaxies are given in Section 4. A discussion in Section 5 and the summary of our result is given in Section 6. We adopt the cosmology $H_{0}=70 km s^{-1} Mpc^{-1}$ and a flat universe where $\Omega_{M}=0.3$ and $\Omega_{\Lambda}=0.7$.

\section{Sample Selection}
The data we use are selected from the quasar catalog \citep{2010AJ....139.2360S} of Sloan Digital Sky Survey data release 7 (SDSS DR7). The SDSS used a dedicated 2.5m wild-field telescope \citep{2006AJ....131.2332G} to image the sky in five broad bands ($u,g,r,i,z$). The QSOs candidates were selected based on their colors \citep{2002AJ....123.2945R} and then observed with fiber-fed double spectrographs with 3$^{\prime\prime}$ diameter fiber which result in getting more emission from host galaxies. . SDSS DR7 quasars catalog contains 105,785 QSOs. In this study, we used the reduced one-dimensional spectral data derived from SDSS DR12 pipeline-processed. The spectra have a wavelength coverage of 3800-9200$\AA$ at a spectral resolution R $\sim$ 1500-2500.

\begin{figure}[h]
	\epsscale{1.0}
	\plotone{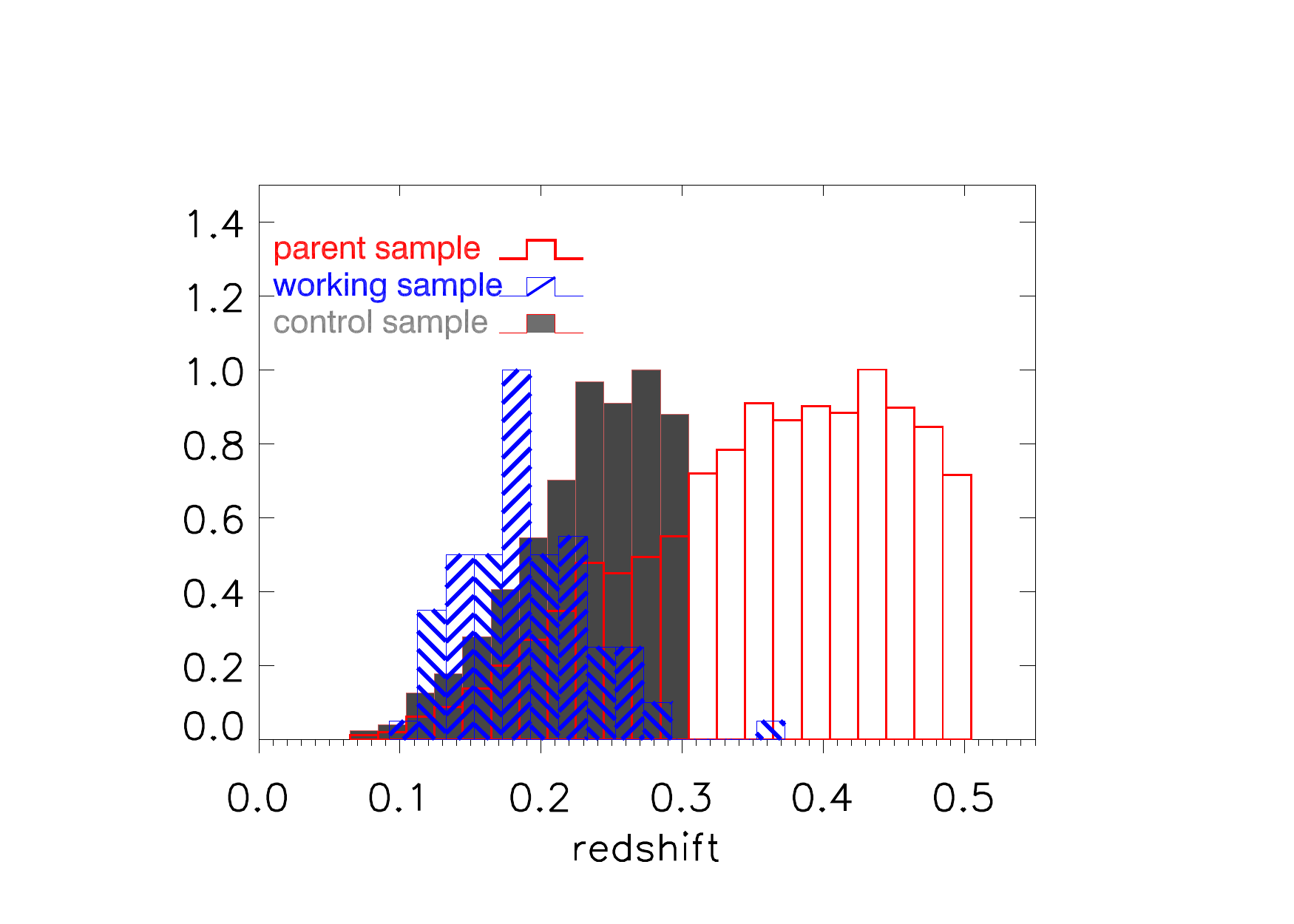}
	\caption{ Redshift distribution for source in the parent sample (red blank), the working sample (blue diagonal) and the control sample (gray filled). The peaks are normalized to one. \label{fig:redshift}}	
\end{figure}

In order to have a  better understanding on the infrared properties of our sources, we built a parent sample by matching SDSS objects with the Wide-field Infrared Survey Explorer (WISE) in 3$^{\prime\prime}$ radius. The survey of WISE covers $95\%$ of sky at 3.4 (W1), 4.6 (W2), 12 (W3), 22 (W4) $\mu m$ with an angular resolution of 6.1$^{\prime\prime}$, 6.4$^{\prime\prime}$, 6.5$^{\prime\prime}$ and 12.0$^{\prime\prime}$ in four bands, achieving, $5\sigma$ point source sensitivities better than 0.08, 0.11, 1, and 6 $mJy$, respectively. We also set the upper limit of the redshift range to $z<0.5$ for two reason: 1) to reject the higher redshift QSOs because they have more luminous AGN which will dilutes the stellar feature; 2) this redshift range allows the spectral to covering the absorption line needed for $STARLIGHT$ analysis. In summary, the parent sample was built as follows: 

1. $z<0.5$,

2. $(S/N)_{WISE}$ of W1, W2, W3 and W4 $> 3$,

3. $(S/N)_{SDSS} \ge 15$,

\noindent where the $(S/N)_{WISE}$ represents the signal-to-noise ratio of photometry in WISE bands and the $(S/N)_{SDSS}$ is the  signal-to-noise ratio of the SDSS spectra. There are total 8490 objects in our parent sample and Figure~\ref{fig:redshift} shows the redshift distribution (red solid line).

Then, we selected the working sample from the parent sample by adding this criterion:

$(S/N)_{stellar} \ge 15$,

\noindent  where the $(S/N)_{stellar}$ is the S/N of stellar composition which were calculated by using synthesis code $STARLIGHT$ and the detail of it will be given in $section 3.1.1$. 

Since the type I QSOs provide great observational challenge owing to their overwhelming brightness of the AGN with respect to the host galaxy, we use above criterion to ensure that all object in the sample have obvious stellar content. At last, there are total 82 objects in the working sample. Table~\ref{table:observation} shows some observational properties of the sources in the working sample. Figure~\ref{fig:redshift} shows the redshift distribution of these source (blue diagonal). The decrease of number of the sources in the working sample in high redshift (only one higher than 0.3) may result from the selection effect that brighter AGNs are more easy to be observed in high redshift and the host galaxy is overwhelmed by these brighter AGN. 

 We also extracted from the parent sample a control sample of QSOs which don't have obvious stellar content. The control sample meeting this criteria:

$(S/N)_{stellar} < 15$.

We also limit the redshift of our control sample to $z < 0.3$. The final control sample is composed of 2183 objects. Figure~\ref{fig:redshift} shows the redshift distribution of the parent sample (red solid line), the working sample (blued diagonal) and the control smaple (gray filled), respectively. We normalize them to 1.0 as peak of each. 

We have to emphasize that the selection method used in this work is different from those preceding ones: 1) This work focuses on the stellar content of type I QSOs which are  brighter than $M_{i} = -22\ mag$. In contrast, many works foucused on low-luminosity type II AGNs \citep{2003MNRAS.346.1055K,2014ApJ...792...84Y,2007ApJ...671.1388D,2003MNRAS.346.1055K,2005MNRAS.356..480T,2000AJ....120.1579Y,2004ApJ...613..109H}.  2) We attempt to study the stellar population of QSOs host which contain the central region of galaxies while some other's work avoid QSOs contamination by off-axis observation (the spectra are obtained with the slit of the spectrograph located a few arcseconds away from the quasar) \citep{2001MNRAS.323..308N,2013ApJ...772..132C}. 3) We don't limit our working sample exclusive to QSOs host with moderate-age stellar population (PSQs), as done by some authors \citep{2000AJ....119...59C,2013ApJ...762...90C,2013ApJ...772...28W}.

\section{Spectral analysis}
\subsection{Spetral Synthesis with Starlight}
 We used the spectral analysis code  $STARLIGHT$ \citep{2005MNRAS.358..363C} to study the stellar population of host galaxies in our working sample. This code searches for the linear combination of $N_{*}$ Simple Stellar Populations (SSP) from evolutionary synthesis models for a best matches of observed spectrum $O_{\lambda}$. The models $M_{\lambda}$ is given by:
 
 $M_{\lambda}=M_{\lambda_{0}}(\sum_{j=1}^{N_{*}}x_{j}b_{j,\lambda}r_{\lambda})\otimes G(v_{\star},\sigma_{\star})$,
 
\noindent where $b_{j,\lambda}$ is the normalized flux of the $j$th SSP at $\lambda_{0}$, $r_{\lambda} \equiv 10^{-0.4(A_{\lambda}-A_{\lambda_{0}})}$ is the reddening term, $M_{\lambda_{0}}$ is the synthetic flux at the normalization wavelength, $x_{j}$ is the population vector and $\otimes$ denotes the convolution operator and G is a Gaussian filter centered at velocity $v_{\star}$ and with dispersion $\sigma_{\star}$. This method carries out the fitting with a mixture of simulated annealing plus Metropolis scheme and Markov Chain Monte Carlo techniques to yield the minimum $\chi^{2}$ value ($\chi^{2} = \sum_{\lambda} [(O_{\lambda}-M_{\lambda})\omega_{\lambda}]$, where $M_{\lambda}$ is the model spectrum and ${\omega_{\lambda}}^{-1}$ is the error in $O_{\lambda}$ at each wavelength bin). The $\chi^{2}/N_{\lambda}$ which we used in $Section 3.1.2.$ is the fit $\chi^{2}$ divided by the number of $\lambda$'s used in the fit.
 
 \subsubsection{Preliminary Fitting}
 Firstly, we used $STARLIGHT$ to fit the integrated spectra for all sources in the parent sample to get the luminosity ratio between the stellar population and AGN. The $STARLIGHT$  is a smarter and fast way of fitting a spectrum and it allows us to contain a power-law. 
 We use SSPs models from \citeauthor{2003MNRAS.344.1000B} (2003, BC03)  (with $N_{\star} = 150$ spectra of 6 metallicaties range from 0.0001 to 0.05 and 25 different ages range from 1Myr to 18Gyr). The reason why we chose BC03 is that it has a wide range of metallicities and it is also widely used in the literature which allowed us to have a comparison between our result and previous works. What's more, the BC03 is the base model for $STARLIGHT$, so it is convenient to use them together. Though the BC03 has a new version (CB07, \cite{2007unpublished}) which includes the new stellar evolution prescription for the TP-AGB evolution., the work  \citep{2013MNRAS.428.1479Z} shown that the BC03 model is still the most successful in reproducing the stellar population of host galaxies.  We also add a power-law spectrum $F_{\lambda} \propto \lambda^{\alpha_{\lambda}}$  which represent the contribution of AGN featureless continuum for preliminary fitting. The power-law spectra index $\alpha$ is -2 which is a traditional value of type 1 QSOs \citep{2007MNRAS.378...83L,2001AJ....122..549V,2011ApJS..194...45S}. The Calzetti law \citep{2000ApJ...533..682C} were used for the reddening during the fitting. We corrected for Galactic extinction using the \cite{1998ApJ...500..525S} maps and extinction curves from \cite{1999PASP..111...63F}.  The input spectra to $STARLIGHT$ contain 4 columns: the wavelength ($\lambda$), the flux ($O_{\lambda}$), the error of flux ($e_{\lambda}$) and the $flag_{\lambda}$ which signals if that pixcel is good or bad. All these message of input spectrum are get from SDSS data release 12.
 
\begin{figure}[h]
	\epsscale{0.65}
	\plotone{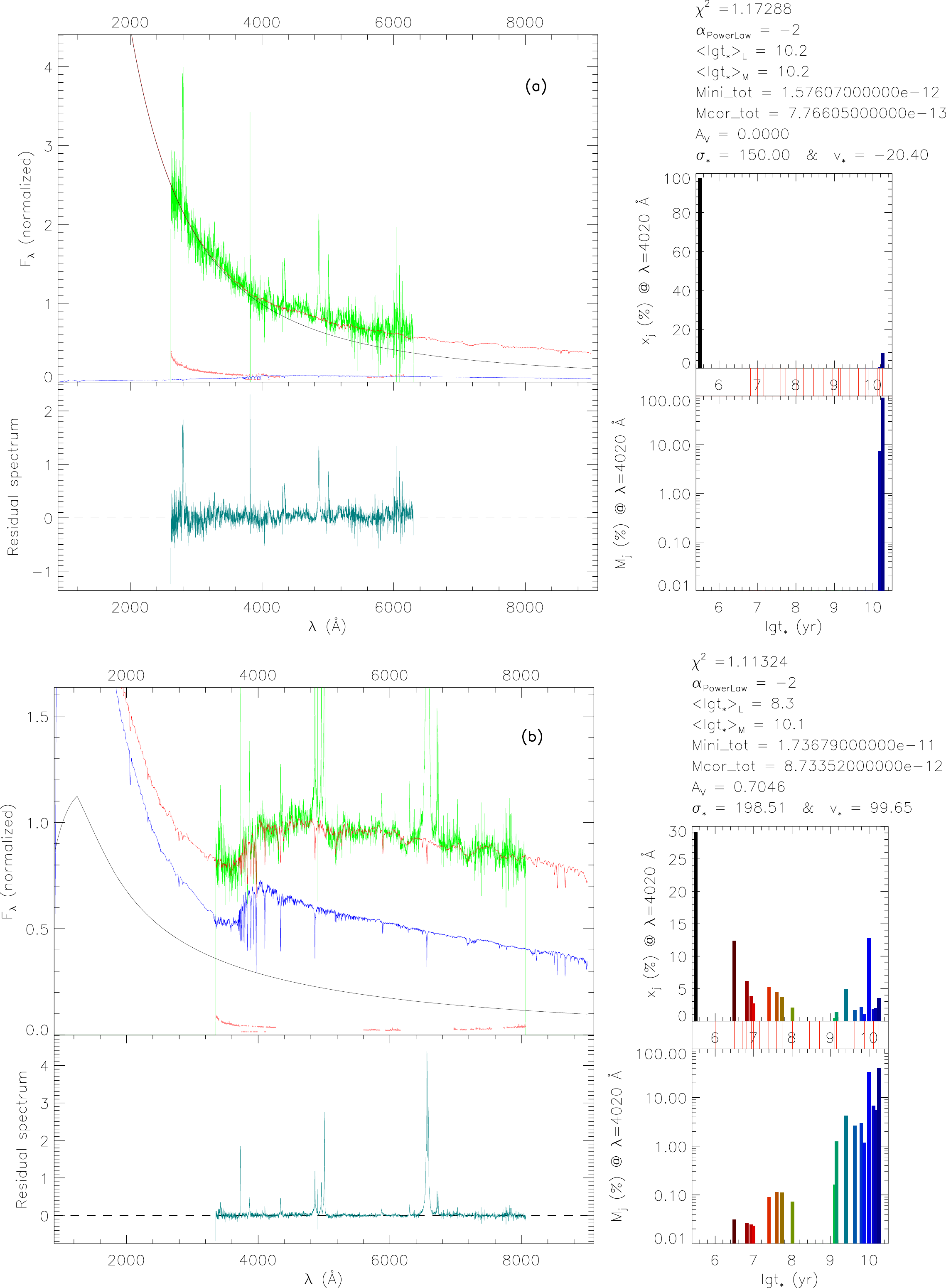}
	\caption{Spectral synthesis of SDSS DR7 QSOs. Panel (a) illustrated QSOs dominated by AGN (SDSS J000011.41+145545.6)---Top left : the observed spectrum $O_{\lambda}$(green), the model spectrum $N_{\lambda}$(red), the host galaxy starlight (blue) ,the power-law (black) and the error spectrum (pink) with the gaps meaning the masked regions and the five times weighted absorption lines. Bottom left: the residual spectrum (dark green). Right: light (top) and mass (bottom) weighted stellar population fraction $x_{j}$ and $\mu_{j}$,respectively. The inserted panel on the right marks the ages of the stellar population templates; Panel (b) illustrated QSOs dominated by stellar population (SDSS J131750.32+601040.8) which has obvious stellar feature such as strong Balmer absorption lines. The meaning of symbols are same as those Figure (a).} 	
	\label{fig:first} 
\end{figure}

   Figure~\ref{fig:first} shows the result of $STARLIGHT$ fitting: the panel (a) is about a normal QSOs from control sample compared with panel (b) representing the objects of the working sample. Figure~\ref{fig:first} (a) shows that most QSOs are dominated by AGN (the luminosity fraction of AGN is almost $100\%$) which can described by a power-law and shows little stellar feature in its spectrum.
 
 In order to gain a sample of QSOs with significant stellar component, we calculate the S/N of host galaxy by
 
 ${(S/N)}_{stellar}={(S/N)}_{SDSS} \cdot \sqrt{1-\eta}$,
 
\noindent where $\eta$ is the luminosity ratio between the AGN and (AGN+host) getting from $STARLIGHT$ fitting; The ${(S/N)}_{stellar}$ is the signal-to-noise ratio of host galaxy. We require the ${(S/N)}_{stellar}$ greater than 15 and the subsample of our sources is total 82 objects.  Figure~\ref{fig:first} (b) show an example in the working sample which are dominated by stellar population with significant stellar feature in its spectrum such as $Ca_{II}$K $\lambda$3933, $Ca_{II}$H $\lambda$3968 and Balmer absorption lines. 

\subsubsection{Formal Fitting and Stellar Populations}
\begin{figure}[h]
	\epsscale{1.0}
	\plotone{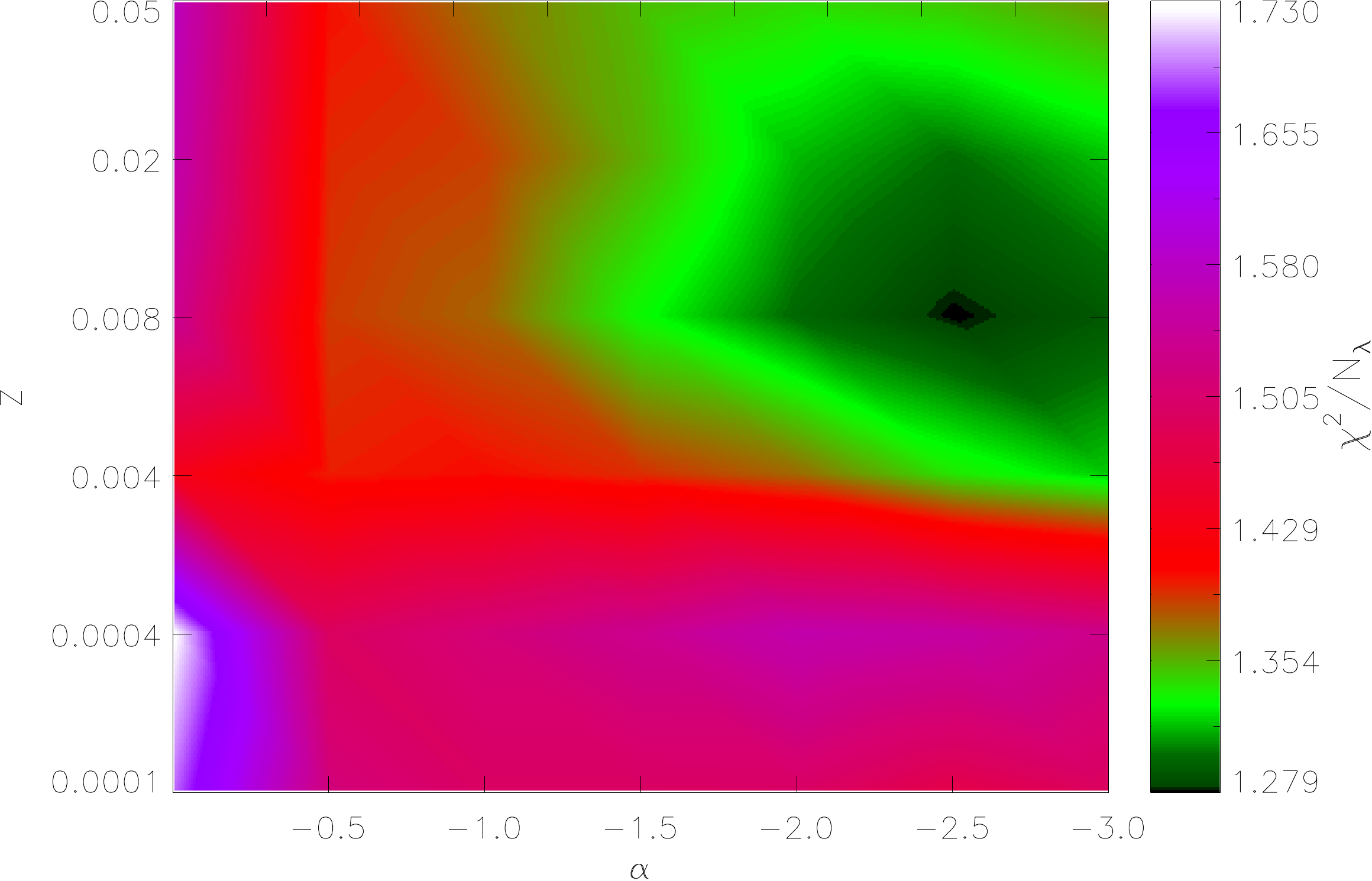}
	\caption{The test fitting of power-law index and metallicity. This figure shows the averaged $\chi^{2}/N_{\lambda}$ changed with metallicities (Z) and power-law indices ($\alpha$).  The color bar is $\chi^{2}/N_{\lambda}$. \label{fig:a_p}}
	
\end{figure}

To make the result more reliable, we get the best-fit metallicity and power-law index by using method proposed by \cite{2010ApJ...718..928M}. We adopted a wide range of AGN power-law slop $\alpha_{\lambda}$ over the optical and UV range from -3.0 to 0 (no power-law fitting with $\alpha_{\lambda}$=0) at intervals of 0.5 to search for the best power-law index. To search for the best-fit metallicity, we carry out a metallicity test with six metallicities ($Z=0.0001, 0.0004, 0.004, 0.008, 0.02, 0.05$) of the BC03 model for each power-law index. Figure~\ref{fig:a_p} shows the test fitting results for one example object with different power-law index $\alpha_{\lambda}$ and six metallicities. We evaluate the fitting quality by the minimum  $\chi^{2}/N_{\lambda}$which is suggested by \cite{2005MNRAS.358..363C}. We averaged it over 100 times fitting with different seed (the random number that need to be appointed during the $STARLIGHT$ fitting). 

\begin{figure}[h]
	\epsscale{1.0}
	\plotone{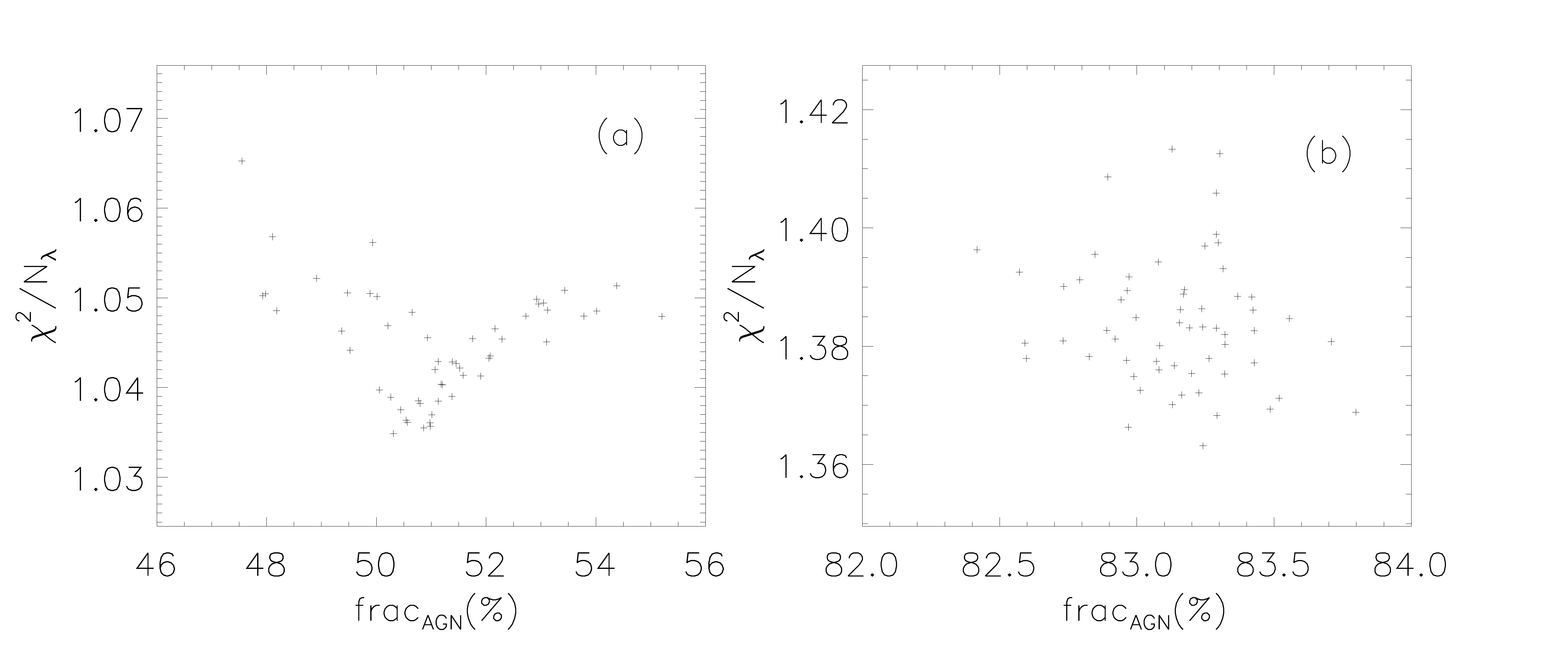}
	\caption{ $\chi^{2}/N_{\lambda}$vs $frac_{AGN}$. The left panel is the example for adopting minimal $\chi^{2}/N_{\lambda}$ as best fitting value; The right panel is the example for adopting averaged $\chi^{2}/N_{\lambda}$ as best fitting value.} 
	\label{fig:chi2s} 
	
\end{figure}

\begin{figure}[h]
	\epsscale{1.0}
	\plotone{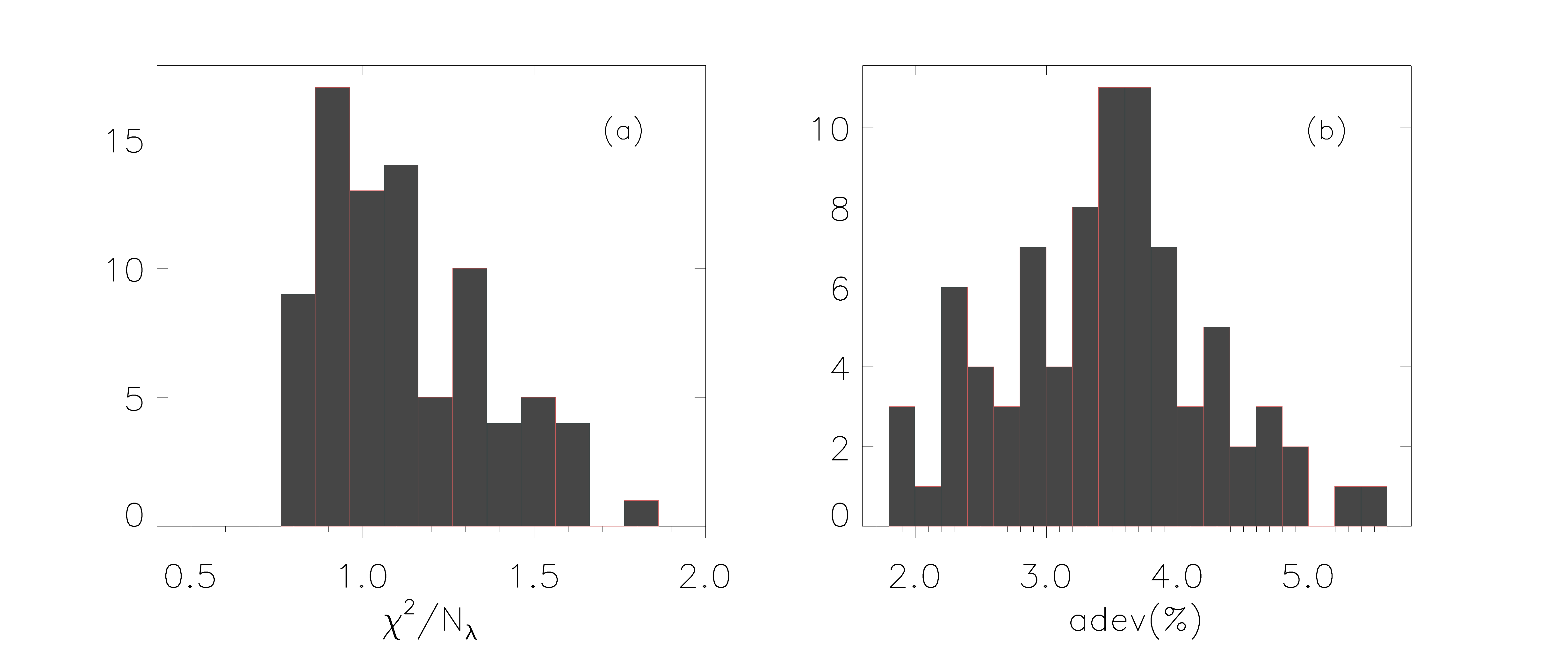}
	\caption{ The distribution of $\chi^{2}/N_{\lambda}$ (panel a) and $adev$ (panel b) , the percentage mean $|O_{\lambda}-M_{\lambda}|/O_{\lambda}$ deviation over all fitted pixels, for 82 sources of the working sample.} 
	\label{fig:adev} 
	
\end{figure}

As for the given metallicity and power law index, there are total 100 fitting results.  Though difference result from random seeds will not change the overall population distribution, a $\sim 10\%$ variation may added to the individual component $x_{j}$ \citep{2010ApJ...718..928M}. In order to get a statistically reliable result, we selected the fitting result of minimal $\chi^{2}/N_{\lambda}$ value (if we have) or mean value over 100 fitting as the best fitting. Figure~\ref{fig:chi2s} shows the distribution of luminosity fraction of AGN ($frac_{AGN}$) and $\chi^{2}/N_{\lambda}$ for two objects in our working sample: we adopt the minimal and mean value in left and right panel, respectively. Besides the $\chi^{2}/N_{\lambda}$ value, we also use the adev vale as an indicator of the quality of fit. The adev gives the percentage mean $|O_{\lambda}-M_{\lambda}|/O_{\lambda}$ deviation over all fitted pixels. The Figure~\ref{fig:adev}  shows the distribution of $\chi^{2}/N_{\lambda}$ value (Panel.a) and adev (Panel.b). Most of sources has a reliable fitting result with $\chi^{2}/N_{\lambda}$ $\sim$ 1. There are ~22\% (18/82) sources have $\chi^{2}/N_{\lambda}  \ge 1.3$ . We find that the main reason accounting for this high $\chi^{2}/N_{\lambda}$ value is related to the fact that some high $frac_{AGN}$ objects are more difficult to fit because of lower stellar content. We have compare the main result getting from the data that contained these 18 objects and the data that do not contain these objects and found that there is no bias between them. So we keep these 18 objects in our work. The adev valus distribution is shown in Figure~\ref{fig:adev} (b) and presents values adev $\lesssim 6$  per cent for all objects, indicating that the model reproduces very well the observed underlying spectra.

\subsection{Classification with SFH}
We calculate the luminosity of the stellar component from UV to optical during the past $\sim$ 1Gyr by reconstructing the UV-to-Optical spectrum. The detailed descriptions of calculation can be found in \cite{2010ApJ...718..928M} and we briefly described the method here.

Because the 3$^{\prime\prime}$ diameter fiber does not cover the whole galaxy, a aperture corrections \citep{2010ApJ...718..928M} are necessary. It is known that the luminosity of the galaxy is dominated by young stellar populations and the AGN, which can both be better traced by the u band, so we adopt the aperture correction for stellar component at the $u$ band derived from

$A=\frac{L_{\star Petro}}{L\star fiber}=\frac{10^{-0.4(m_{petro}-m_{AGN})-1}}{10^{-0.4(m_{fiber}-m_{AGN})-1}}$.

 We rebuild the rest-frame model spectrum $F_{i}(\lambda,t)$ of stellar with whole UV-to-Optical wavelengths coverage ($912 \sim 9000 \AA$) corrected for extinction by

$F_{i}(\lambda,t)=Mcor\_tot \sum_{j=1}^{N}\frac{\mu_{j}}{f_{\star,j,t_{0}}}f_{\star,j,t}B_{\lambda,j,t}$,

\noindent where $F_{i}(\lambda,t)$ is the stellar spectrum (corrected for extinction) at a given time $t$. $t$ can be the time in the past or at the present (equal $t_{0}$). $Mcor\_tot$ is present stellar mass obtained from the spectral synthesis after aperture correction, $\mu_{j}$ is mass-weighted fraction, $B_{\lambda,j,t}$ are BC03 SSP templates without normalization, $f_{\star,j,t_{0}}$ is the present ($t_{0}$) fraction of remaining stellar mass to the initial mass of population $j$, $f_{\star,j,t}$ is such fraction at a given time $t$. We estimate the UV-to-Optical luminosity of the past 25, 100, 290, 500 and 900 Myr separately.

we calculate the UV-to-Optical luminosity history of the host galaxies by

$L_{UV\_Optical,t}=\int_{912}^{9000}[F_{i}(\lambda,t)]d\lambda$ .

 To test the feasibility of our method, we also reconstructed the spectrum in present time ($t_{0}$) by adding the dust extinction and double-index power. The formula is
 
$F_0(\lambda,t_{0})=[L_{UV\_Optical,0}+F_p(\lambda)]\times10^{-0.4(A_\lambda-A_V)}$,

\noindent where $F_p(\lambda)$ is a double power-law spectrum of AGN. The spectral indexes we used here are given by $\alpha=-1$ for $\lambda<1250 \AA$ \citep{2008MNRAS.386.1252H} and $\alpha$ given by starlight for $\lambda>1250 \AA$. The $A_V$ is obtained from the spectral synthesis. Figure~\ref{fig:history} shows the reconstructed model spectrum (red solid line) superimposed by the observed one (green solid line). This model spectrum is used as $F_0(\lambda,t_{0})$.

 At last, we give a simple classification of our working sample based on the SFH ($L_{UV\_Optical,t}$) of the host galaxy. In summary, there are two main features of these SFH: one is a dramatically enhanced star formation about 900 Myr ago, and moderate one recently (with in 500 Myr). Since their prototypes are unknown, we focus on these two features and attempt to classify our source according to their SFH in our work. At last, we classify our working sample into 4 types: 
 
 (1)$type O$ : Don't have obvious star formation activity in the past 900 Myr. The star formation activity in this objects could took place 1 Gyr ago. There are total 16 sources in this type.
 
 (2)$type A$ : Only have moderate star formation activity within 500Myr. There are total 20 sources in this type;
 
 (3)$type B$ : Only have dramatically enhanced star formation activity about 900 Myr ago. There are total 11 sources in this type;
 
 (4)$type AB$ : Have both two main feature of SFH ( dramatically enhanced one and moderate one). There are total 35 sources in this type.
 
\begin{figure}[h]
	\epsscale{1.0}
	\plotone{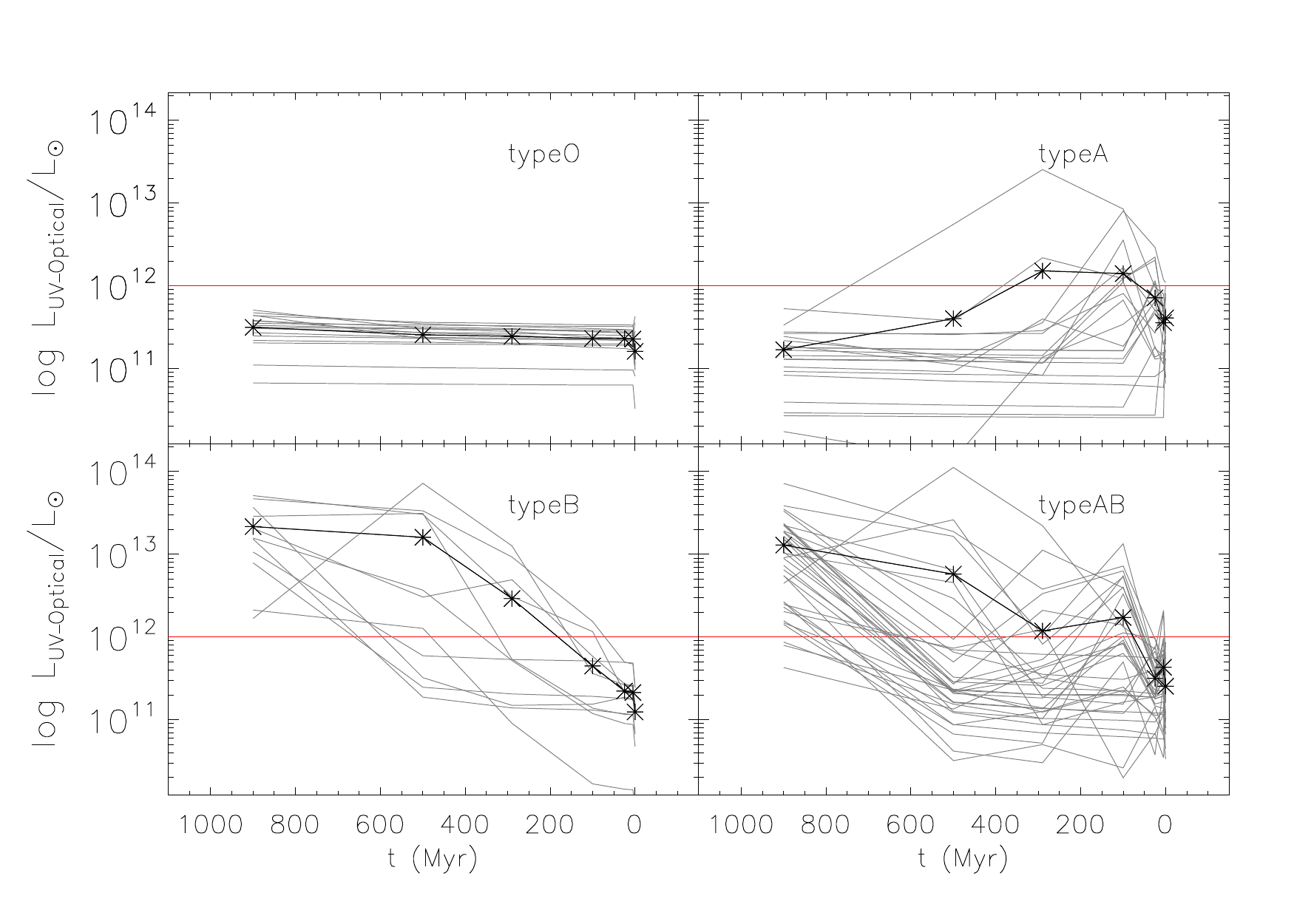}
	\caption{The four panels display the averaged $L_{UV\_Optical}$ (912-9000 $\AA$) of source of the working sample of four different types: $type O$, $type A$, $type B$ and $type AB$. The gray lines represent the individual objects for corresponding classes. \label{fig:classfy}}
\end{figure}

Figure~\ref{fig:classfy}  shows the mean star formation histories of these four different type (dark line) which is added with individual objects ( gray lines ) for corresponding classes.

\section{Result \& Analysis}
\subsection{Composite Spectra}
In order to characterize the spectrum-to-spectrum difference for these four SFH types, we make a combination for each type by normalizing each individual spectrum at 5100$\AA$ and then computing the average value of $F_{\lambda}$ in bins of $\lambda$. 

\begin{figure}[h]
	\epsscale{1.0}
	\plotone{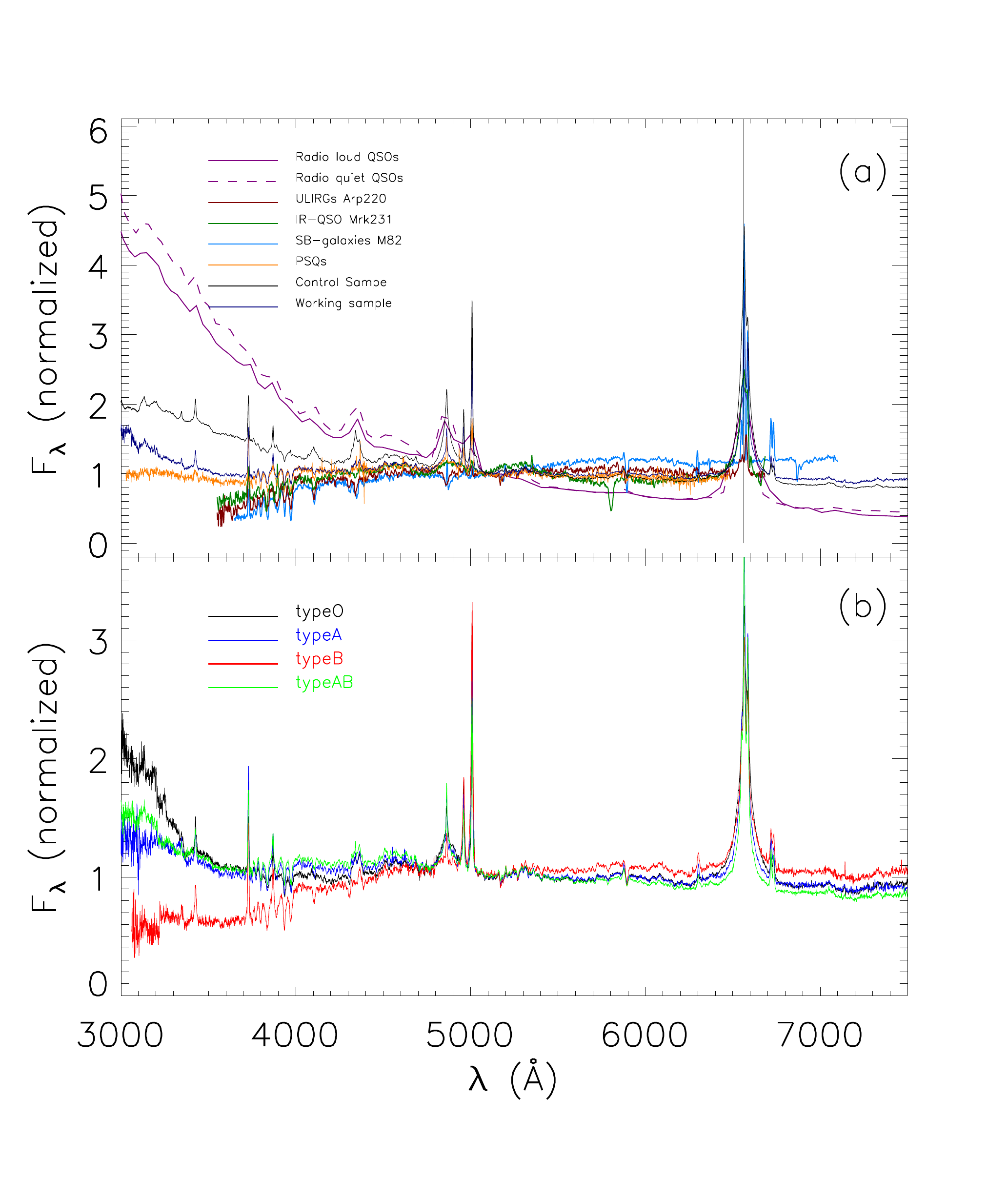}
	\caption{ The panel (a) shows the composite spectra of the radio loud QSOs (solid purple line), radio quiet QSOs (dash purple line), PSQs (orange line), IR QSOs (dark green line), ULIRGs (dark red line), starburst galaxy (cyan blue line) and the control sample (black line); The panel (b) shows the composite spectra of: $type O$ (black), $type A$  (blue), $type B$ (red) and $type AB$ (green).} 
	\label{fig:com_spec} 
\end{figure}

Figure~\ref{fig:com_spec} shows the composites spectral of sources sample in the working . For more explicit, we divide the spectrum into two panels. In the panel (a), we compare composite spectrum of the souses in the working sample (blue solid line) with those of others. For example, we plot the QSOs spectra from \cite{2011ApJS..196....2S}, which presented the SEDs of 85 optically bright, non-blazar QSOs (27 radio-quiet and 58 radio-loud) over the wavelength from radio to X-ray. The purple solid line represents the radio-loud QSOs and the purple dashed line represents the radio-quiet QSOs. Additionally, we also show the spectra of others in Figure~\ref{fig:com_spec}: the black solid line represents the control sample, the orange solid line represents the PSQ from \cite{2011ApJ...741..106C}, the dark green solid line represents the Mrk 231 (IR-QSOs, \citealt{2006ApJS..164...81M}), the dark red solid line represents the Arp 220 (ULIRGs, \citealt{2006ApJS..164...81M}) and the cyan blue solid line represents the M82 (SB galaxiy, \citealt{1992ApJS...79..255K}). When compared to control sample, the working sample are more luminous in the red (wavelength longer than 5100$\AA$) and closed to PSQs which indicating that the sources in the working sample have significant contribution from stellar content and this conclusion is consistent with the result of \cite{2015MNRAS.449.2374C} that PSQs are overall red compared to typical QSOs' color with a significant contribution from a post-starburst stellar population. It is clear from the panel (a) in Figure~\ref{fig:com_spec} that the slope of spectra of the sources in the working sample is intermediate between those of  IR-QSOs and QSOs, which imply a possibility that our objects could be in the evolutionary stage from IR-QSOs to typical optical QSOs.

 Panel (b) of Figure~\ref{fig:com_spec} gives a comparison for our four types. The black, blue, red, and green spectral represents the $type O$, $type A$, $type B$ and $type AB$, respectively. The $type A$ and $type AB$ have significant Balmer absorption lines such as $H\delta$ along with steeper continuum in the blue, which may be attributed to young stellar population or AGN activity. If considering there is no significant high AGN fraction (see later in $Section\ 4.2$ ) in $type A$ and $type AB$, we suggest that the steeper continuum in the blue is result from young stellar population. The $type O$ has even steeper continuum in the blue. it has invisible Balmer absorption lines and significant high AGN fraction, which means that the AGN emission is the major contribution in this type. The continuum of $type B$ are much flatter and the $Ca_{II}$K $\lambda$3933, $Ca_{II}$H $\lambda$3968 absorption lines are also obvious, which suggest this types are hosted in galaxies with significant contribution from older stellar content and not dominated by AGN.

\begin{figure}[h]
	\epsscale{1.0}
	\plotone{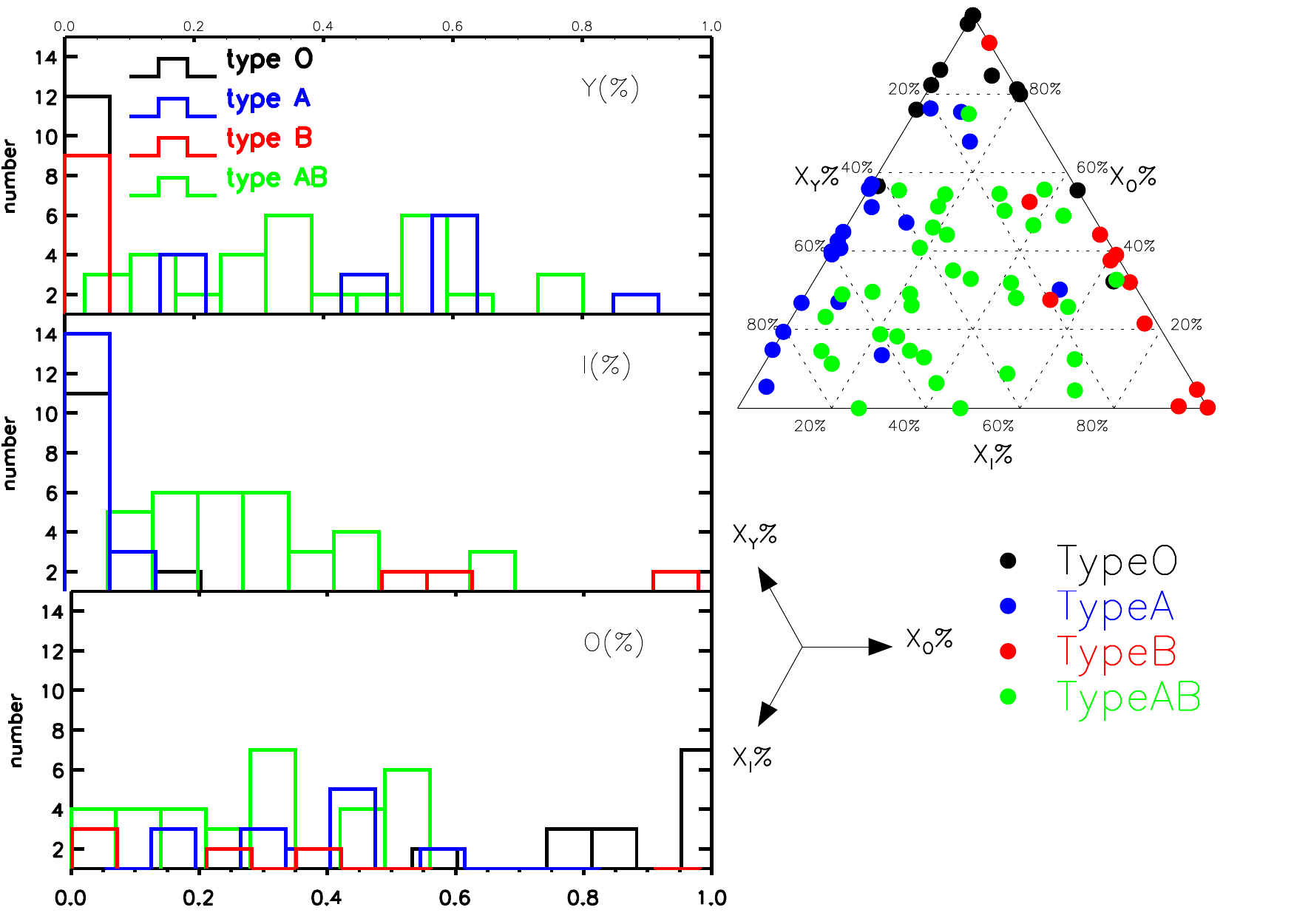}
	\caption{The age bin description was plotted in a trigonometric coordinate panel (right panel). black filled circles denote $type O$, blue filled circles denote the $type A$, red filled circles denote the $type B$ and green filled circles denote the $type AB$. The arrows in picture illustrate the direction to readout the $X_{Y}\%$, $X_{I}\%$, $X_{O}\%$, respectively. The same result are also depicted in the form of histograms (left panel). The black, blue, red and green histograms in each figure represent the $type O$, $type A$, $typeB$ and $typeAB$, respectively.\label{fig:YIO}}	
\end{figure}

 As discussed in \cite{2004MNRAS.355..273C}, individual stellar population components are very uncertain because the existence of multiple solutions in stellar population and the further binning of the age will give a coarse but more robust description of the star formation history (SFH). We separate the stellar population into three components as suggested by \cite{2004MNRAS.355..273C}: ``young" ($X_{Y}$,$t \le 1.0\times 10^{8}yr$), ``intermediate" ($X_{I}$,$1.6\times 1.0^{8}yr\le t \le 1.27\times 10^{9}yr$) and ``old" ($X_{O}$,$t \ge 1.43\times 10^9 yr$). The diversity among four types objects is show in Figure~\ref{fig:YIO}, where we present a trigonometric coordinate of $X_{Y}+X_{I}+X_{O}=1$ panel. It is clear that the $type O$ (black dot) is dominated by old age stellar population while the $type A$ (blue dot) is dominated by young stars. $typr B$ (red dot) have a major contribution from intermediate population and the $type AB$ (green dot) have mixed contribution from both young-aged and intermediate-aged stellar population, which is consistent with the result of composite spectra. To be more clearly, we depict the same result in the form of histograms (for each stellar population). The black, blue, red and green histograms in each figure represent the $type O$, $type A$, $typeB$ and $typeAB$, respectively.

\subsection{AGN Luminosity Fraction and WISE Color}
\begin{figure}[h]
	\epsscale{1.0}
	\plotone{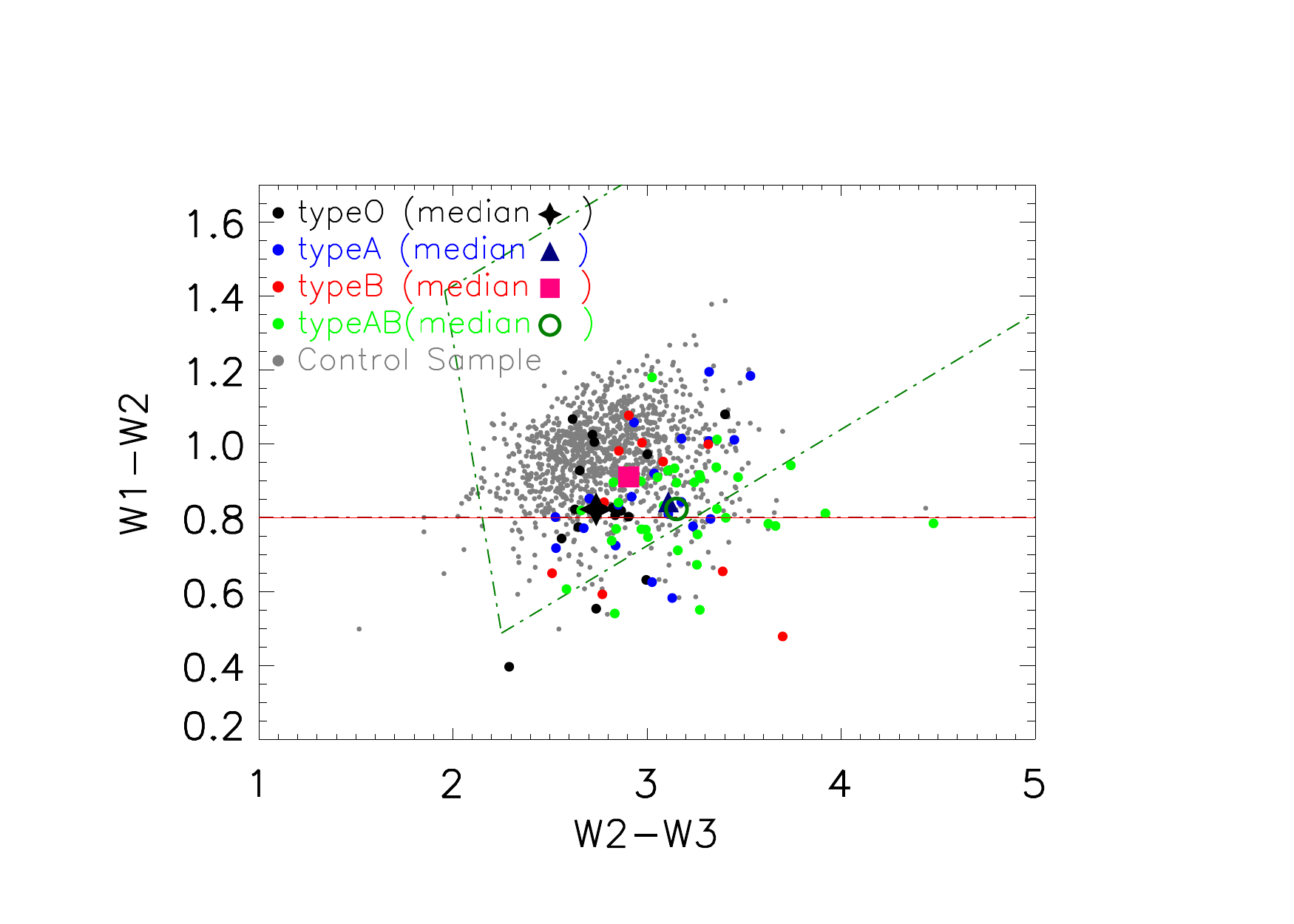}
	\caption{WISE colors of our working sample, which shows the $W1-W2\ vs\ W2-W3$: the meaning of the symbols are same as those in Figure~\ref{fig:YIO}  and the median value of each type are represented by different symbols (typeO: black star, typeA: blue taiangle, typeB: red square, typeAB: green open circle). There are different characteristic region used to select AGN. AGN wedge (dark green dot-dashed line) was defined by \cite{2012MNRAS.426.3271M} and the mid-IR criteria (dark red dot-dashed line) was proposed by \cite{2012ApJ...753...30S}. }
	\label{fig:WISE}
\end{figure}

The WISE has provided the data in the near- and mid-infrared.  \cite{2012ApJ...753...30S} presented a simple mid-IR color criterion (W1-W2 $\ge$ 0.8) to identify AGN. Figure~\ref{fig:WISE} shows the distribution of $W1-W2\ vs\ W2-W3$ for our 82 objects and the control sample. The median value of each type are also represented by different symbols (typeO: black star, typeA: blue taiangle, typeB: red square, typeAB: green open circle). As expected, most control sample have W1-W2 $\ge$ 0.8 (dark red solid line) while some objects of our working sample have W1-W2 bluer than 0.8. \cite{2012ApJ...753...30S} showed that a bluer W1-W2 is caused by host galaxy contamination in $z < 2$. Additionally, the dark green dot-dashed line illustrates the selection of AGN using W1, W2 and W3 \citep{2012MNRAS.426.3271M}. Not surprisingly, the sources in our working sample lie around AGN boundary with redder W2-W3 and bluer W1-W2. We give a Kolmogorov-Smirnov (KS) test between the working sample and control sample (Table~\ref{table:KS_WISE}). The result shows that the probabilities that the working sample and the control sample are drawn from the same distribution are $P_{KS} \ll 0.001$. The composite AGN/galaxy SED provided by \cite{2012MNRAS.426.3271M} also suggested that the blend with host galaxy will lead the objects lie out the AGN wedge. Moreover, their work also show that the galaxy with old stellar content has W2-W3 color bluer than that of star-formation, which is consistent with our result: the W2-W3 color of $type O$ and $type B$ tend to be bluer than $type A$ and $type AB$ in color-color diagram. So we give a KS-test between $type O+type B$ and $type A+type AB$  in W23 and the resulting probability, $P_{KS} \ll 0.001$, suggests that they are come from different distribution.

\begin{figure}[h]
	\epsscale{1.0}
	\plotone{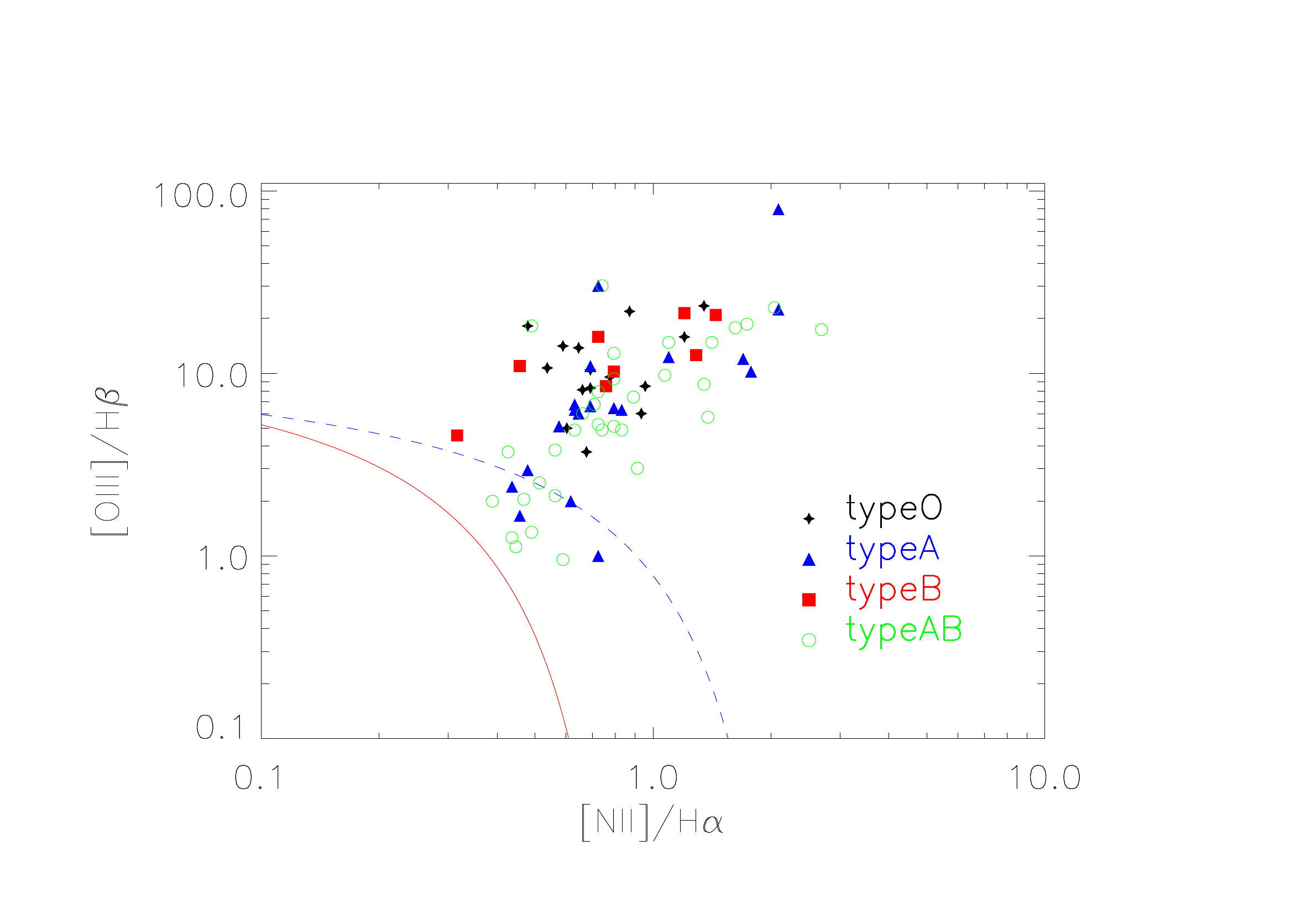}
	\caption{$[NII]/H{\alpha}$ versus $[OIII]/H{\beta}$ diagnostic diagram with line ratios and samples classified according to the combined classification scheme of \cite{2001ApJ...556..121K} (blue dashed line) and \cite{2003MNRAS.346.1055K} (red solid line). Each type are represented by different symbols (typeO: black star, typeA: blue triangle, typeB: red square, typeAB: green open circle). \label{fig:BPT}}	
\end{figure}

The $[NII]/H\alpha$ versus $[OIII]/H\beta$ diagnostic diagram (BPT diagram) is commonly used to  separate star formation from AGN activity \citep{1981PASP...93....5B,1987ApJS...63..295V}. The AGN sequence branches from the enriched end of the star-forming sequence and moves towards larger $[N_{II}]/H_{\alpha}$ and $[O_{III}]/H_{\beta}$ ratios as the AGN fraction increases. \cite{2007ApJ...668...87W} defined a quantity $d_{AGN}$ which measures the distance of galaxies from \cite{2001ApJ...556..121K} theoretical upper bound of pure star formation, along lines parallel to the AGN sequence. \cite{2014MNRAS.444.3961D} also calculated relative AGN fractions by populating the composite region of BPT diagram with starburst-AGN mixing model. With the help of EW of narrow emission line from \cite{2011ApJS..194...45S}, we plot our working sample in the BPT diagram (Figure~\ref{fig:BPT}). We observe a starburst-AGN mixing sequence of the working sample except $type O$ exclusively occupy the region with high $frac_{AGN}$. 

\subsection{Correlation with AGN Properties}
The tight correlation between the black hole and the bulge within which it resides ($M_{BH}\ vs\ M_{bulge}$, $L_{BH}\ vs\ L_{bulge}$, $M_{BH}\ vs\ \sigma_{bulge}$) reveals a close connection between black holes and their host galaxies. \cite{2004ApJ...613..109H} found that in the most present-day accretion occurs onto black hole with masses less than $10^8 M_{\odot}$ and young stellar population. In the study of (sub)mm-loud QSOs, \cite{2008ChJAA...8...12H} found a trend that the star formation rate increases with the accretion rate. They also found the star formation rate decrease with the central black hole mass and suggested that the higher Eddington ratios of IR-QSOs imply that they are in the evolution stage toward QSOs. \cite{2004ApJ...613..109H} found the similar result, at low redshift more massive galaxies tend to have older stellar population.

\cite{2011ApJS..194...45S} presented a compilation of properties of SDSS DR7 quasar catalog. In this product, they compiled continuum and emission measurements, as well as other quantities such as virial black hole mass and Eddington ratio estimates. With the help of these quasar properties, the expected correlations between AGN properties and stellar population are indeed found.

\begin{figure}[h]
	\epsscale{1.0}
	\plotone{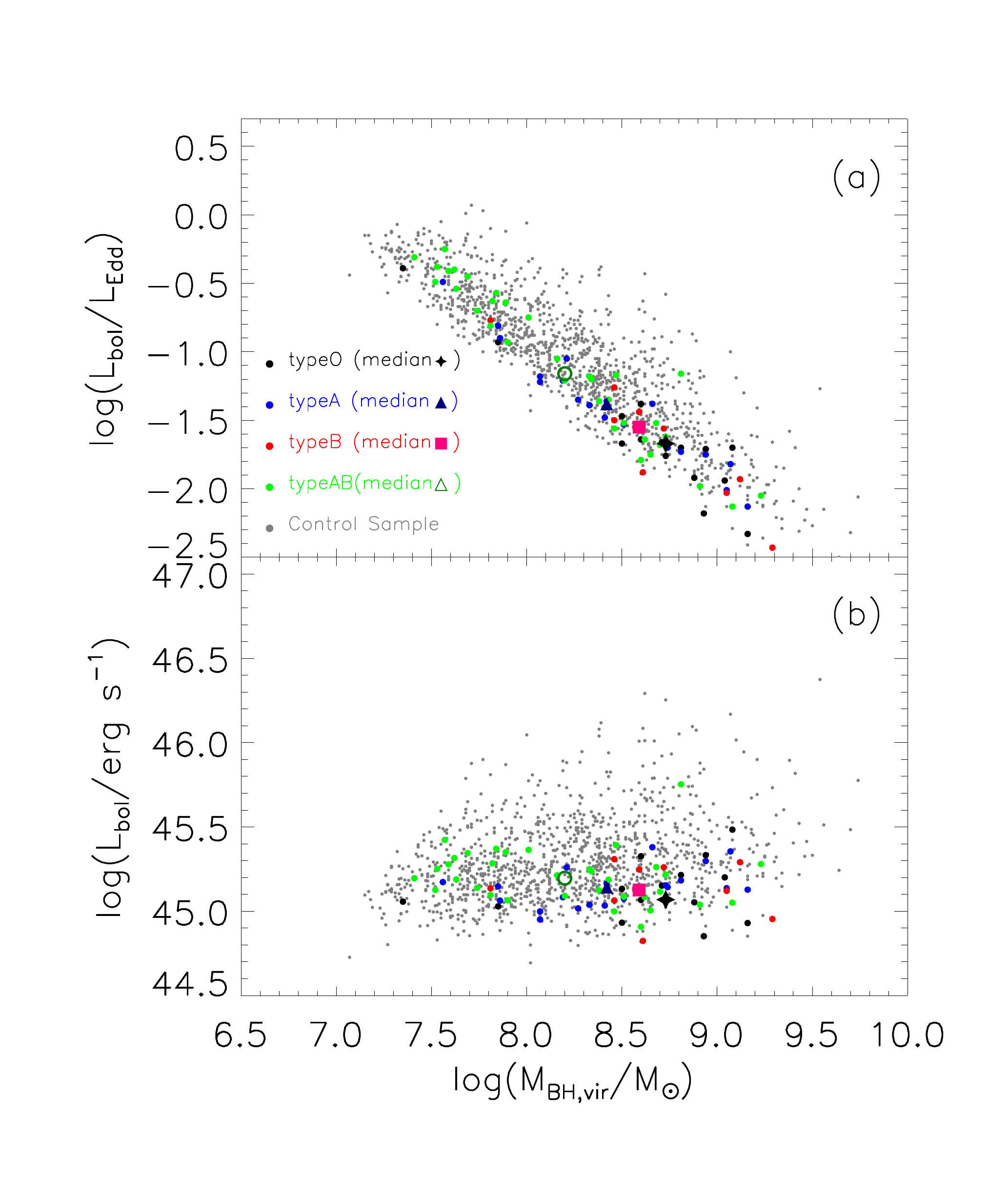}
	\caption{Panel (a) in Figure~\ref{fig:BH} shows the relation between $M_{BH,vir}$ and $L_{bol}/L_{Edd}$. The meaning of symbols are same as those in Figure~\ref{fig:WISE}. The panel (b) shows the relation between $M_{BH,vir}$ and $L_{bol}$.}
	\label{fig:BH}  
\end{figure}

\begin{figure} 
	\centering 
	\subfigure[]{ \label{fig:subfig:a} 
		\includegraphics[width=3.0in]{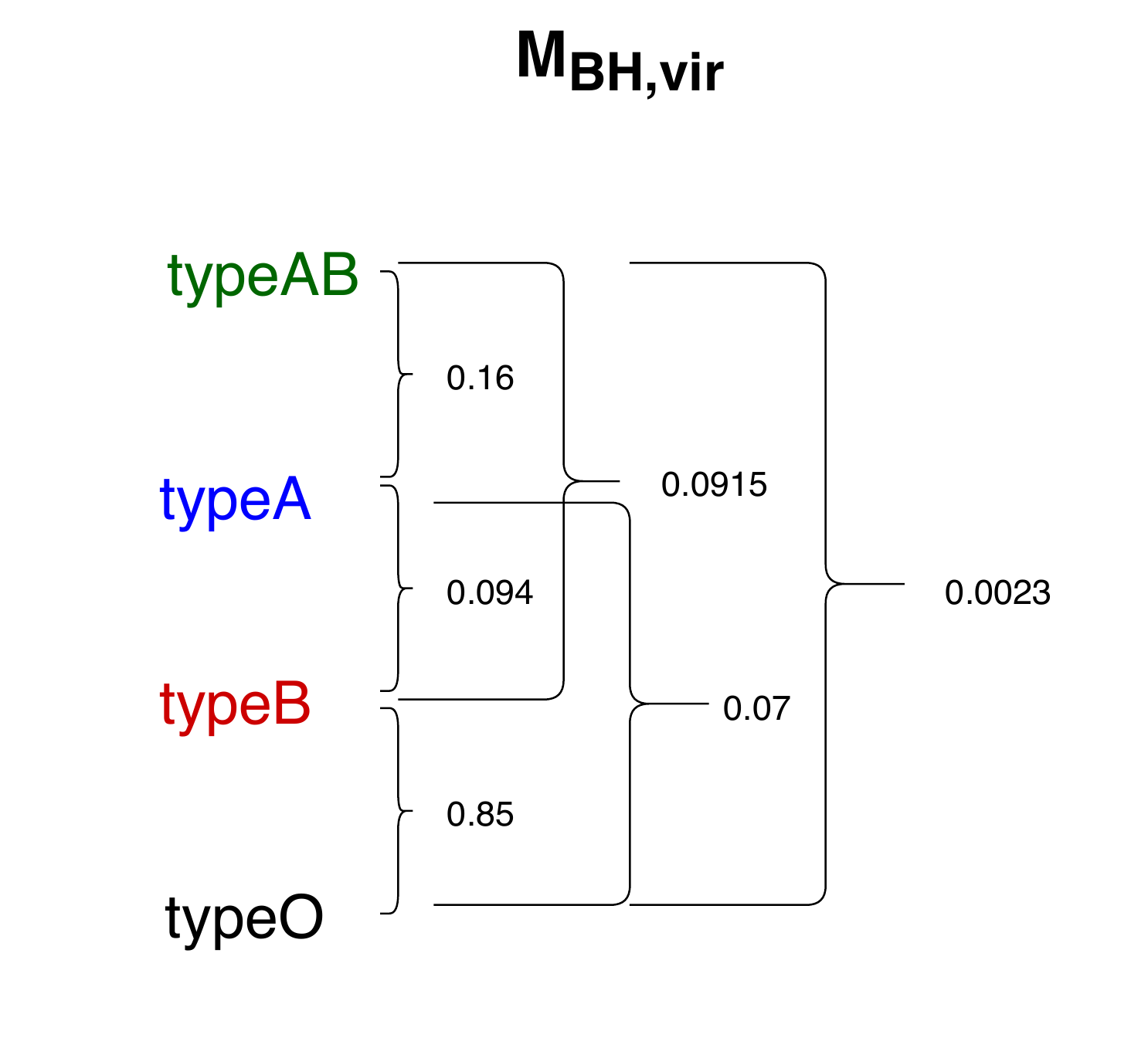}} 
	\hspace{0.02in}
	\subfigure[]{ \label{fig:subfig:b} 
		\includegraphics[width=3.0in]{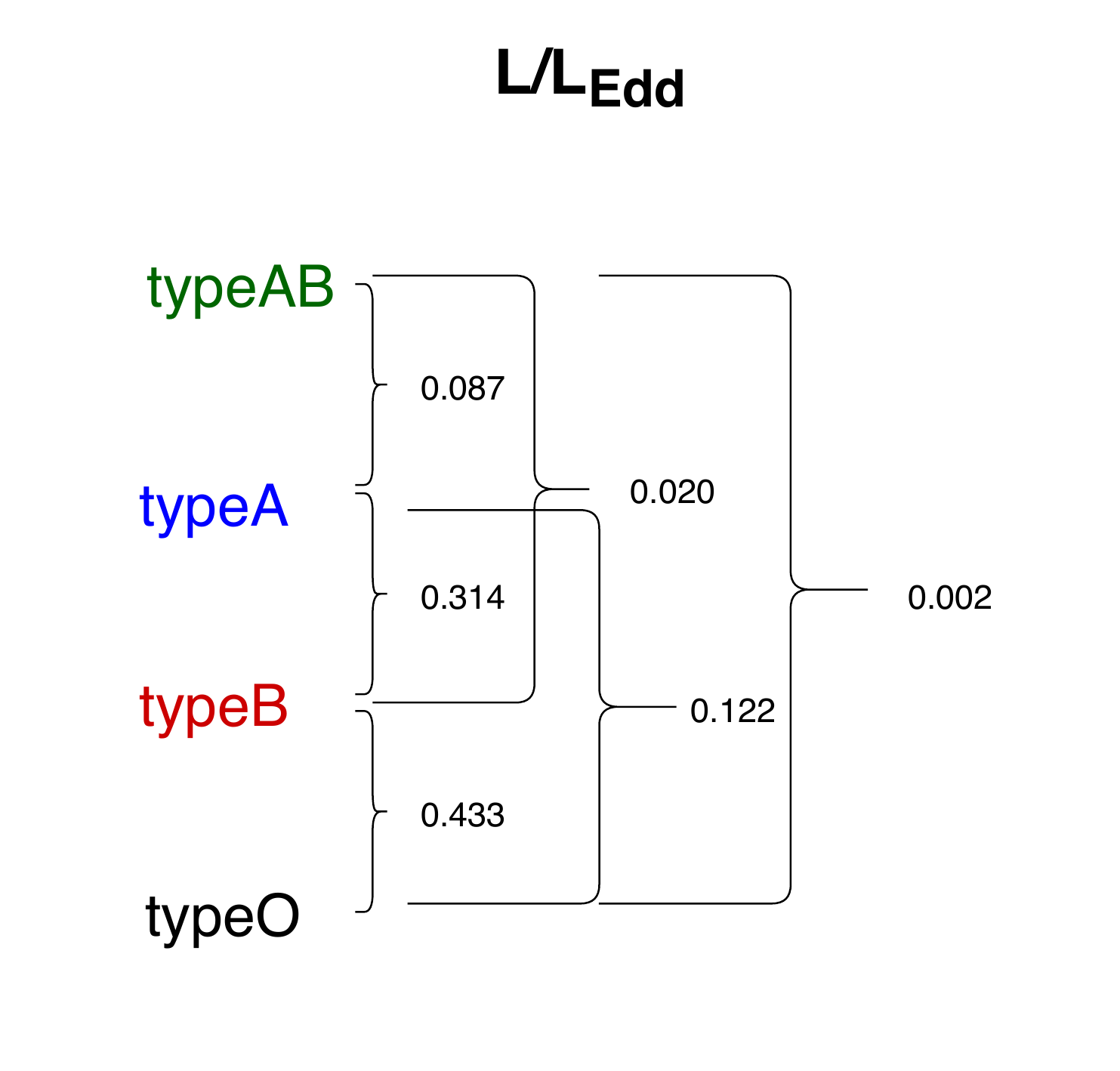}} 
	\caption{ The result of KS-test for $M_{BH}$ and $L/L_{Edd}$: the probability of p value between any two types.} 
	\label{fig:KS_AGN1} 
\end{figure}

Panel (a) of Figure~\ref{fig:BH} shows a strong correlation between black hole mass and Eddington ratio ($M_{BH}$ increases as $L/L_{Edd}$ decreases) for the sources the working sample and control sample (gray dot). By applying the Pearson test, the statistical significance of this two variables is 0.001 and the correlation coefficient is 0.998, which means this two variables are related.  Our result indicate that the low-mass black holes are more active than massive ones. In contrast, the more massive black holes are currently experiencing less additional accretion. The blue triangle, red square, green open circle and black star in Figure~\ref{fig:BH} denote the median value of $typeA$, $typeB$, $typeAB$ and $typeO$, respectively. By comparing these median value, we find that the QSOs with previous star formation activity ($typeA$, $typeB$ and $typeAB$) tend to have high Eddington ratio (stronger black hole active),  statistically. Alternatively, the QSOs which have both former and recent star formation ($typeAB$) tend to have stronger AGN activity, while the $typeO$ which have no obvious star formation activity in the past are the inactive ones. Generally speaking, there may be correlation between the star formation and black hole activity, which may imply a co-evolution between SFH of host galaxies and AGN activity.  The KS-test of both $M_{BH}$ and $L/L_{Edd}$ show that the working sample and the control sample are draw from the different distribution with $P_{KS}$ $\ll 0.001$. We also use the KS-test to test the significant difference of $M_{BH}$ and $L/L_{Edd}$ among these four types and the result are shown in Figure~\ref{fig:KS_AGN1} (a) and (b), respectively. The number next to the braces gives the $P_{KS}$ of this two types. Combining with Figure~\ref{fig:BH}, we can see that if the different between the median value of two types is greater, then the difference significance between these two type will also be greater (with small $P_{KS}$).  The $P_{KS}$ between $typeAB$ (most active among four type) and $typeO$ (most inactive among four type) show that this two type are draw from the different distribution both in $M_{BH}$ and $L/L_{Edd}$. 

Panel (b) of Figure~\ref{fig:BH} shows a relationship between black holes mass and luminosity of the working sample.  We quantify the Eddington and BH masses by calculating their mean values and standard errors (SE) for all four types and the control sample (Table~\ref{table:mean_SE}). Compared with control QSOs, our working sample seems have lower Eddington ratio as well as lower luminosity which may be result from the selection effect (the high luminosity QSOs tend to have a more powerful AGN and may overwhelm the light of host galaxies). The KS-test of luminosity show that the working sample and the control sample are draw from the different distribution with $P_{KS}$  $\ll 0.001$.

\subsection{Interaction of host galaxies}
\begin{figure}[h]
	\epsscale{1.0}
	\plotone{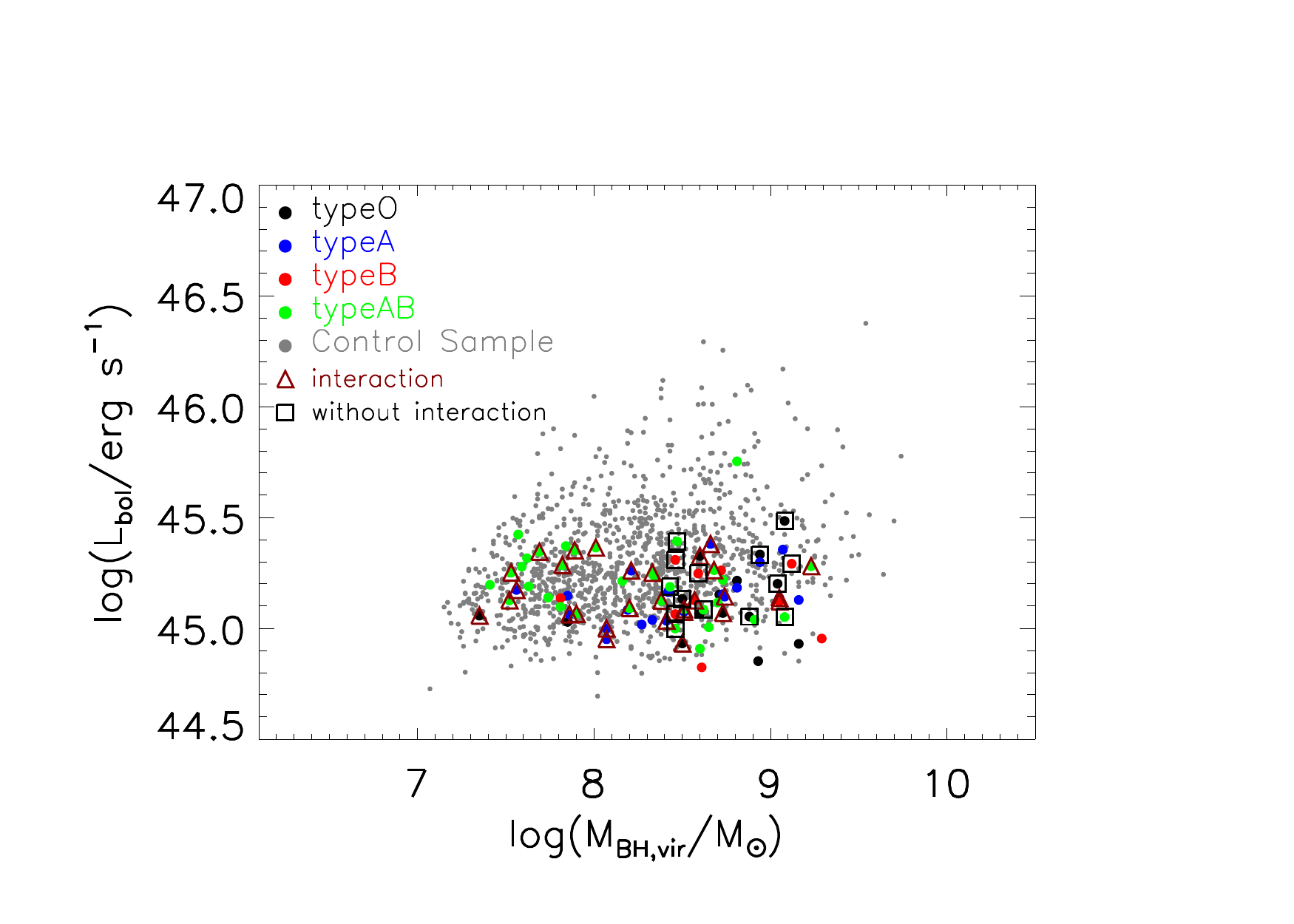}
	\caption{The interaction condition of host galaxies were ploted on the program of $M_{BH}\ vs\ L_{bol}$. The black squares represent the host galaxies without interaction while the red triangles represent the host galaxies with obvious tidal feature. The meaning of other symbols are same as those in previous figure. \label{fig:IM2}}	
\end{figure}

A long term discussion concerning the properties of the galaxies hosting QSOs is its morphology. To study the host galaxy morphology properties and their interaction, we use images from the SDSS. The main challenge in understanding the host galaxy is the poor spatial resolution of the ground-based observations and combined with the bright nuclei hindered the nature of the QSOs host. We just classify our sources into tow types: Y(30) with obvious interactivity, which was represented with red triangle in Figure~\ref{fig:IM2} and N (14) which don't shows any tidal feature and no companion (black squares). The rest 38 source is hard to be classified by SDSS image. It is interesting that the host galaxies without interaction are exclusive locate in high black hole mass and alliance with $type O$ and $type B$ (9/14). This result may indicate that objects without recent star formation and high black hole mass may already relax from the interactivation and evolve to the quiescent ellipse galaxies.

\section{Discussion}
By studying the stellar population of host galaxies in type I QSOs among different SFH types, we find that the stellar population is associated with the AGN physical properties. These result is consistent with the finding of \cite{1988ApJ...325...74S} and \cite{2000ApJ...539L..13G} , at least some QSOs are in the advanced merge stage.

\subsection{Spectral comparison between our working sample and others}
 \cite{1988ApJ...328L..35S} suggested a evolutionary sequence from ULIRGs to QSOs.  \cite{2005ApJ...625...78H} and \cite{2008MNRAS.390..336C} also proposed that at last some IR-QSOs are at a transitional stage from ULIRGs to classical QSOs. The work of \cite{2012ApJ...758L..39T} showed that most luminous AGN (QSOs, $M_{R} \le -22$) seem to be triggered by major mergers. If so, the composite spectra of the working sample will be consistent with the idea that they are hybrids of AGN and starbursts (or post-starbust ). We compare different kind of spectra in Figure~\ref{fig:com_spec}. As we can see in panel (a), the spectra of QSOs from \cite{2011ApJS..196....2S} (both radio quiet and radio loud) are brighter in the wavelength shorter than 5100$\AA$ and even much brighter than control sample. It may because most objects selected from \cite{2011ApJS..196....2S} are UV-bright-AGN. The ULIRGs have roughly similar SED to the starburst galaxies while IR-QSOs is more bright in the shorter wavelength, which are consistent with the evolutionary mentioned before. Additionally, the spectra of our working sample is closed to PSQs from \cite{2011ApJ...741..106C}, which means they have similar $frac_{AGN}$ or even in the similar evolution stage. The study of \cite{2011ApJ...741..106C} shows that the PSQs have a starburst within ~100 Myr which is smaller than our age range. Combined with the analysis in $Section\ 4.1$, our objects may in the evolutionary stage from IR-QSOs to typical optical QSOs.
 
 \subsection{The AGN properties in the evolution}
 In former analysis, we find a compelling correlation between the star-busts (both former dramatically enhanced and recent moderate star formation) and AGN properties. Considering the typical QSOs' lifetime are expected to be $\sim 10^8$ yr, the dramatically enhanced star formation can't have been triggered at the same time as they occurred few hundred Myr ago. However, there are some theories can explain the correlation between black hole activity and former star formation \citep{2000AJ....120.1750C,2002AJ....123..225W,2003AJ....126...63H,2004ApJ...600..580G,2006ApJS..163...50H,2007MNRAS.378...83L,2008MNRAS.384..386C,2012MNRAS.420L...8H}.
  
 The previous work suggested that star-burst may occur at the early stage of merger or interaction, when the AGN has not been triggered yet and then followed by decrease in accretion of AGN with the aging of stellar content \citep{2000AJ....120.1750C,2004ApJ...600..580G,2006ApJS..163...50H}. \cite{2012MNRAS.420L...8H} showed that such a time delay can occur for purely dynamical reasons. His simulations showed firstly the gas move toward center and gives rise to the star formation. Then, the gas flowing further inwards by losing angular momentum and produces a time delay between star formation and AGN activity. What's more, many numerical simulations show that the star formation as well as AGN activity is episodic \citep{2008ApJS..175..356H,2012ApJ...748L...7V,2012ApJ...746..108T}, which depend on the detail of the merge (orbits, morphological type of progenitors). In this picture, the former dramatically enhanced star formation of QSO hosts in the working sample are induced in the early stages of galaxies merger. As the galaxies continue to merge for next few hundred Myr, the gas flowing further inwards central regions, followed by the AGN activity. The recent moderate star formation in our working sample may be triggered by a minor merge due the accretion of a satellite \citep{2008MNRAS.384..386C}. Many works \citep{2002AJ....123..225W,2003AJ....126...63H,2007MNRAS.378...83L} suggest that minor merger that do not produce dramatically enhanced star formation, while still fuelling the AGN. Our result is consistent with that of the former work. \cite{2007ApJ...671.1388D} analyzed the star formation in the nuclei of nine Seyfert galaxies, which also show a possible starbursts in the last 10-300 Myr. In the work of \cite{2009ApJ...692L..19S}, the AGN (obscured and unobscured) appear to be prevalent in the "green valley" on the color-magnitude diagram. They suggested that there is a ~100Myr time delay between the shutdown of star formation and detectable AGN. The research of \cite{2004ApJ...613..109H} also used type 2 AGNs to investigate the accretion-driven growth of super-massive black holes and found that bulge formation and black hole formation are tightly coupled in present-day. The study of \cite{2012A&A...540A.109S,2013MNRAS.429....2F} shows a higher SFRs of AGN host galaxies than that of inactive galaxies with the same stellar mass and redshift.
 
\section{Summary}
We have studies the stellar population of 82 host galaxies of type I QSOs selected from the cross-matched SDSS DR7 QSOs and WISE catalog with the S/N of stellar content great than 15. Our method is a powerful technique for selecting the type I QSOs with obvious host component and investigating the properties of their host galaxies. Compared with WISE color and BPT diagram, we have shown that this technique can efficiently separate the spectroscopic components and give a reliably stellar content and $frac_{AGN}$. Furthermore, we can also classify our working sample into four type by SFH ($type O$, $type A$, $type B$ and $type AB$) of host with this method.

The composite spectra, age distribution and WISE color-color diagram show that:(1) the stellar population have a significant contribution to the observed emission in our sources; (2) The $type O$ are dominated by older stellar content; $type A$ are dominated by young stellar content; the $type B$ are dominated by intermediate stellar content; $type AB$ have a mixed component of young and intermediate stellar content.

Considering the AGN properties such as black hole mass and Eddington ratio, we suggest that there is a co-evolution between AGN and host galaxies. Our result also shows there is a time delay between the peak of star formation and black hole accretion and then both of them decrease slowly. In addition, the host galaxies that do not show sign of interactive ( no tidal feature and companion galaxies ) are exclusive reside in the object with big black hole mass and relative old stellar population and imply they are the object relaxed form interactive. Most of our sources may be in the transition phase from IR-QSOs to classical optical QSOs.

\section*{Acknowledgments}
We grateful to the referee and editors for their careful reading of the paper and useful comments and suggestions. We also thank the people, Chao-jian Wu, Fang Yang and Wei Du, for useful assist that improved the work.  This project is supported by National Key R\&D Program of China(No.2017YFA0402704),
and the National Natural Science Foundation of China (Grant Nos. 11733006,
11225316, 11173030 and U1531245). This project is also supported by the China Ministry of Science and Technology under the State Key Development Program for Basic Research ,2014CB845705,（LAMOST973）.

This work is partially Supported by the Open Project Program of the Key Laboratory of Optical
Astronomy, National Astronomical Observatories, Chinese Academy of Sciences. The author thank the useful SDSS database and the DR7 edition of the SDSS quasars Catalog. Funding for the SDSS has been provided by the Alfred P. Sloan Foundation, the Participating Institutions, the National Science Foundation, the U.S. Department of Energy,
the National Aeronautics and Space Administration, the Japanese Monbukagakusho, the
Max Planck Society, and the Higher Education Funding Council for England. This work also makes use of data products from the Wide-field Infrared Survey Explorer, which is a joint project of the University of California, Los Angeles, and the Jet Propulsion Laboratory/California Institute of Technology, funded by the National Aeronautics and Space Administration. 

\begin{thebibliography}{99}
	
	\bibitem[Aird et al.(2010)]{2010MNRAS.401.2531A} Aird, J., Nandra, K., Laird, E.~S., et al.\ 2010, \mnras, 401, 2531 
	\bibitem[Assef et al.(2013)]{2013ApJ...772...26A} Assef, R.~J., Stern, D., Kochanek, C.~S., et al.\ 2013, \apj, 772, 26
	\bibitem[Baldwin et al.(1981)]{1981PASP...93....5B} Baldwin, J.~A., Phillips, M.~M., \& Terlevich, R.\ 1981, \pasp, 93, 5
	\bibitem[Barnes \& Hernquist(1991)]{1991ApJ...370L..65B} Barnes, J.~E., \& Hernquist, L.~E.\ 1991, \apjl, 370, L65
	\bibitem[Barnes \& Hernquist(1992)]{1992ARA&A..30..705B} Barnes, J.~E., \& Hernquist, L.\ 1992, \araa, 30, 705
	\bibitem[Barnes \& Hernquist(1996)]{1996ApJ...471..115B} Barnes, J.~E., \& Hernquist, L.\ 1996, \apj, 471, 115
	\bibitem[Barth et al.(2008)]{2008ApJ...683L.119B} Barth, A.~J., Bentz, M.~C., Greene, J.~E., \& Ho, L.~C.\ 2008, \apjl, 683, L119 
	\bibitem[Blank \& Duschl(2016)]{2016MNRAS.462.2246B} Blank, M., \& Duschl, W.~J.\ 2016, \mnras, 462, 2246 
	\bibitem[Bruzual \& Charlot(2003)]{2003MNRAS.344.1000B} Bruzual, G., \& Charlot, S.\ 2003, \mnras, 344, 1000
	\bibitem[Caccianiga et al.(2015)]{2015MNRAS.451.1795C} Caccianiga, A., Ant{\'o}n, S., Ballo, L., et al.\ 2015, \mnras, 451, 1795
	\bibitem[Cales et al.(2011)]{2011ApJ...741..106C} Cales, S.~L., Brotherton, M.~S., Shang, Z., et al.\ 2011, \apj, 741, 106 
	\bibitem[Cales et al.(2013)]{2013ApJ...762...90C} Cales, S.~L., Brotherton, M.~S., Shang, Z., et al.\ 2013, \apj, 762, 90
	\bibitem[Cales \& Brotherton(2015)]{2015MNRAS.449.2374C} Cales, S.~L., \& Brotherton, M.~S.\ 2015, \mnras, 449, 2374	
	\bibitem[Calzetti et al.(2000)]{2000ApJ...533..682C} Calzetti, D., Armus, L., Bohlin, R.~C., et al.\ 2000, \apj, 533, 682 
	\bibitem[Canalizo et al.(2000)]{2000AJ....119...59C} Canalizo, G., Stockton, A., Brotherton, M.~S., \& van Breugel, W.\ 2000, \aj, 119, 59
	\bibitem[Canalizo \& Stockton(2000)]{2000AJ....120.1750C} Canalizo, G., \& Stockton, A.\ 2000, \aj, 120, 1750
	\bibitem[Canalizo \& Stockton(2013)]{2013ApJ...772..132C} Canalizo, G., \& Stockton, A.\ 2013, \apj, 772, 132
	\bibitem[Canalizo \& Stockton(2001)]{2001ApJ...555..719C} Canalizo, G., \& Stockton, A.\ 2001, \apj, 555, 719
	\bibitem[Cao et al.(2006)]{2006ChJAA...6..197C} Cao, C., Wu, H., Wang, J.-L., et al.\ 2006, \cjaa, 6, 197
	\bibitem[Cao et al.(2008)]{2008MNRAS.390..336C} Cao, C., Xia, X.~Y., Wu, H., et al.\ 2008, \mnras, 390, 336 
	\bibitem[Charlot \& Bruzual(2007)]{2007unpublished} Charlot, S. \& Bruzual, G. 2007,models distributed on demand (CB07)
	\bibitem[Cid Fernandes et al.(2004)]{2004MNRAS.355..273C} Cid Fernandes, R., Gu, Q., Melnick, J., et al.\ 2004, \mnras, 355, 273
	\bibitem[Cid Fernandes et al.(2005)]{2005MNRAS.358..363C} Cid Fernandes, R., Mateus, A., Sodr{\'e}, L., Stasi{\'n}ska, G., \& Gomes, J.~M.\ 2005, \mnras, 358, 363 
	\bibitem[Clements et al.(1996)]{1996MNRAS.279..477C} Clements, D.~L., Sutherland, W.~J., McMahon, R.~G., \& Saunders, W.\ 1996, \mnras, 279, 477 
	\bibitem[Cox et al.(2008)]{2008MNRAS.384..386C} Cox, T.~J., Jonsson, P., Somerville, R.~S., Primack, J.~R., \& Dekel, A.\ 2008, \mnras, 384, 386 
	\bibitem[Cui et al.(2001)]{2001AJ....122...63C} Cui, J., Xia, X.-Y., Deng, Z.-G., Mao, S., \& Zou, Z.-L.\ 2001, \aj, 122, 63
	\bibitem[Davies et al.(2007)]{2007ApJ...671.1388D} Davies, R.~I., M{\"u}ller S{\'a}nchez, F., Genzel, R., et al.\ 2007, \apj, 671, 1388 
	\bibitem[Davies et al.(2014)]{2014MNRAS.444.3961D} Davies, R.~L., Kewley, L.~J., Ho, I.-T., \& Dopita, M.~A.\ 2014, \mnras, 444, 3961
	\bibitem[Di Matteo et al.(2005)]{2005Natur.433..604D} Di Matteo, T., Springel, V., \& Hernquist, L.\ 2005, \nat, 433, 604
	\bibitem[Ferrarese \& Merritt(2000)]{2000ApJ...539L...9F} Ferrarese, L., \& Merritt, D.\ 2000, \apjl, 539, L9 
	\bibitem[Jahnke et al.(2007)]{2007MNRAS.378...23J} Jahnke, K., Wisotzki, L., Courbin, F., \& Letawe, G.\ 2007, \mnras, 378, 23
	\bibitem[Kauffmann et al.(2003)]{2003MNRAS.346.1055K} Kauffmann, G., Heckman, T.~M., Tremonti, C., et al.\ 2003, \mnras, 346, 1055
	\bibitem[Kewley et al.(2001)]{2001ApJ...556..121K} Kewley, L.~J., Dopita, M.~A., Sutherland, R.~S., Heisler, C.~A., \& Trevena, J.\ 2001, \apj, 556, 121
	\bibitem[Kennicutt(1992)]{1992ApJS...79..255K} Kennicutt, R.~C., Jr.\ 1992, \apjs, 79, 255 
	\bibitem[Kim et al.(2002)]{2002ApJS..143..277K} Kim, D.-C., Veilleux, S., \& Sanders, D.~B.\ 2002, \apjs, 143, 277 \
	\bibitem[Kormendy \& Richstone(1995)]{1995ARA&A..33..581K} Kormendy, J., \& Richstone, D.\ 1995, \araa, 33, 581
	\bibitem[Kormendy \& Gebhardt(2001)]{2001AIPC..586..363K} Kormendy, J., \& Gebhardt, K.\ 2001, 20th Texas Symposium on relativistic astrophysics, 586, 363 
	\bibitem[Kormendy \& Sanders(1992)]{1992ApJ...390L..53K} Kormendy, J., \& Sanders, D.~B.\ 1992, \apjl, 390, L53  
	\bibitem[Dunlop et al.(2003)]{2003MNRAS.340.1095D} Dunlop, J.~S., McLure, R.~J., Kukula, M.~J., et al.\ 2003, \mnras, 340, 1095
	\bibitem[Engel et al.(2010)]{2010A&A...524A..56E} Engel, H., Davies, R.~I., Genzel, R., et al.\ 2010, \aap, 524, A56
	\bibitem[Farrah et al.(2001)]{2001MNRAS.326.1333F} Farrah, D., Rowan-Robinson, M., Oliver, S., et al.\ 2001, \mnras, 326, 1333
	\bibitem[Fitzpatrick(1999)]{1999PASP..111...63F} Fitzpatrick, E.~L.\ 1999, \pasp, 111, 63 
	\bibitem[Ferrarese \& Merritt(2000)]{2000ApJ...539L...9F} Ferrarese, L., \& Merritt, D.\ 2000, \apjl, 539, L9 
	\bibitem[Floyd et al.(2013)]{2013MNRAS.429....2F} Floyd, D.~J.~E., Dunlop, J.~S., Kukula, M.~J., et al.\ 2013, \mnras, 429, 2 
	\bibitem[Gebhardt et al.(2000)]{2000ApJ...539L..13G} Gebhardt, K., Bender, R., Bower, G., et al.\ 2000, \apjl, 539, L13 
	\bibitem[Gebhardt et al.(2000)]{2000ApJ...543L...5G} Gebhardt, K., Kormendy, J., Ho, L.~C., et al.\ 2000, \apjl, 543, L5 
	\bibitem[Granato et al.(2004)]{2004ApJ...600..580G} Granato, G.~L., De Zotti, G., Silva, L., Bressan, A., \& Danese, L.\ 2004, \apj, 600, 580 
	\bibitem[Gunn et al.(2006)]{2006AJ....131.2332G} Gunn, J.~E., Siegmund, W.~A., Mannery, E.~J., et al.\ 2006, \aj, 131, 2332 
	\bibitem[Hao et al.(2005)]{2005ApJ...625...78H} Hao, C.~N., Xia, X.~Y., Mao, S., Wu, H., \& Deng, Z.~G.\ 2005, \apj, 625, 78
	\bibitem[Hao et al.(2008)]{2008ChJAA...8...12H} Hao, C.-N., Xia, X.-Y., Shu-DeMao, Deng, Z.-G., \& Wu, H.\ 2008, \cjaa, 8, 12 
	\bibitem[Hatziminaoglou et al.(2008)]{2008MNRAS.386.1252H} Hatziminaoglou, E., Fritz, J., Franceschini, A., et al.\ 2008, \mnras, 386, 1252
	\bibitem[Heckman et al.(2004)]{2004ApJ...613..109H} Heckman, T.~M., Kauffmann, G., Brinchmann, J., et al.\ 2004, \apj, 613, 109 
	\bibitem[Hopkins(2004)]{2004ApJ...615..209H} Hopkins, A.~M.\ 2004, \apj, 615, 209
	\bibitem[Hopkins et al.(2006)]{2006ApJS..163...50H} Hopkins, P.~F., Hernquist, L., Cox, T.~J., Robertson, B., \& Springel, V.\ 2006, \apjs, 163, 50
	\bibitem[Hopkins et al.(2008)]{2008ApJS..175..356H} Hopkins, P.~F., Hernquist, L., Cox, T.~J., \& Kere{\v s}, D.\ 2008, \apjs, 175, 356-389 
	\bibitem[Hopkins(2012)]{2012MNRAS.420L...8H} Hopkins, P.~F.\ 2012, \mnras, 420, L8 
	\bibitem[Hutchings et al.(2003)]{2003AJ....126...63H} Hutchings, J.~B., Maddox, N., Cutri, R.~M., \& Nelson, B.~O.\ 2003, \aj, 126, 63
	\bibitem[Larson \& Tinsley(1978)]{1978ApJ...219...46L} Larson, R.~B., \& Tinsley, B.~M.\ 1978, \apj, 219, 46 
	\bibitem[Letawe et al.(2007)]{2007MNRAS.378...83L} Letawe, G., Magain, P., Courbin, F., et al.\ 2007, \mnras, 378, 82
	\bibitem[Maraston(2005)]{2005MNRAS.362..799M} Maraston, C.\ 2005, \mnras, 362, 799
	\bibitem[Mateos et al.(2012)]{2012MNRAS.426.3271M} Mateos, S., Alonso-Herrero, A., Carrera, F.~J., et al.\ 2012, \mnras, 426, 3271 
	\bibitem[Magain et al.(1998)]{1998ApJ...494..472M} Magain, P., Courbin, F., \& Sohy, S.\ 1998, \apj, 494, 472 
	\bibitem[Magorrian et al.(1998)]{1998AJ....115.2285M} Magorrian, J., Tremaine, S., Richstone, D., et al.\ 1998, \aj, 115, 2285 
	\bibitem[McLure et al.(1999)]{1999MNRAS.308..377M} McLure, R.~J., Kukula, M.~J., Dunlop, J.~S., et al.\ 1999, \mnras, 308, 377
	\bibitem[Meng et al.(2010)]{2010ApJ...718..928M} Meng, X.-M., Wu, H., Gu, Q.-S., Wang, J., \& Cao, C.\ 2010, \apj, 718, 928 
	\bibitem[Meng et al.(2011)]{2011RAA....11..419M} Meng, X.-M., Wu, H., \& Cao, C.\ 2011, Research in Astronomy and Astrophysics, 11, 419
	\bibitem[Merritt \& Ferrarese(2001)]{2001MNRAS.320L..30M} Merritt, D., \& Ferrarese, L.\ 2001, \mnras, 320, L30
	\bibitem[Moustakas \& Kennicutt(2006)]{2006ApJS..164...81M} Moustakas, J., \& Kennicutt, R.~C., Jr.\ 2006, \apjs, 164, 81 
	\bibitem[Murphy et al.(1996)]{1996AJ....111.1025M} Murphy, T.~W., Jr., Armus, L., Matthews, K., et al.\ 1996, \aj, 111, 1025
	\bibitem[Nolan et al.(2001)]{2001MNRAS.323..308N} Nolan, L.~A., Dunlop, J.~S., Kukula, M.~J., et al.\ 2001, \mnras, 323, 308
	\bibitem[Rembold et al.(2017)]{2017MNRAS.472.4382R} Rembold, S.~B., Shimoia, J.~S., Storchi-Bergmann, T., et al.\ 2017, \mnras, 472, 4382
	\bibitem[Richards et al.(2002)]{2002AJ....123.2945R} Richards, G.~T., Fan, X., Newberg, H.~J., et al.\ 2002, \aj, 123, 2945
	\bibitem[Rieke \& Low(1972)]{1972ApJ...176L..95R} Rieke, G.~H., \& Low, F.~J.\ 1972, \apjl, 176, L95
	\bibitem[Rieke \& Lebofsky(1979)]{1979ARA&A..17..477R} Rieke, G.~H., \& Lebofsky, M.~J.\ 1979, \araa, 17, 477
	\bibitem[S{\'a}nchez et al.(2004)]{2004ApJ...614..586S} S{\'a}nchez, S.~F., Jahnke, K., Wisotzki, L., et al.\ 2004, \apj, 614, 586
	\bibitem[S{\'a}nchez et al.(2018)]{2018RMxAA..54..217S} S{\'a}nchez, S.~F., Avila-Reese, V., Hernandez-Toledo, H., et al.\ 2018, \rmxaa, 54, 217 
	\bibitem[Sanders \& Mirabel(1996)]{1996ARA&A..34..749S} Sanders, D.~B., \& Mirabel, I.~F.\ 1996, \araa, 34, 749 
	\bibitem[Sanders et al.(1988)]{1988ApJ...325...74S} Sanders, D.~B., Soifer, B.~T., Elias, J.~H., et al.\ 1988, \apj, 325, 74  
	\bibitem[Sanders et al.(1988)]{1988ApJ...328L..35S} Sanders, D.~B., Soifer, B.~T., Elias, J.~H., Neugebauer, G., \& Matthews, K.\ 1988, \apjl, 328, L35 
	\bibitem[Shang et al.(2011)]{2011ApJS..196....2S} Shang, Z., Brotherton, M.~S., Wills, B.~J., et al.\ 2011, \apjs, 196, 2 
	\bibitem[Santini et al.(2012)]{2012A&A...540A.109S} Santini, P., Rosario, D.~J., Shao, L., et al.\ 2012, \aap, 540, A109 
	\bibitem[Schawinski et al.(2009)]{2009ApJ...692L..19S} Schawinski, K., Virani, S., Simmons, B., et al.\ 2009, \apjl, 692, L19 
	\bibitem[Schlegel et al.(1998)]{1998ApJ...500..525S} Schlegel, D.~J., Finkbeiner, D.~P., \& Davis, M.\ 1998, \apj, 500, 525
	\bibitem[Schneider et al.(2010)]{2010AJ....139.2360S} Schneider, D.~P., Richards, G.~T., Hall, P.~B., et al.\ 2010, \aj, 139, 2360
	\bibitem[Shen et al.(2011)]{2011ApJS..194...45S} Shen, Y., Richards, G.~T., Strauss, M.~A., et al.\ 2011, \apjs, 194, 45 
	\bibitem[Silverman et al.(2008)]{2008ApJ...679..118S} Silverman, J.~D., Green, P.~J., Barkhouse, W.~A., et al.\ 2008, \apj, 679, 118-139 
	\bibitem[Springel et al.(2005)]{2005MNRAS.361..776S} Springel, V., Di Matteo, T., \& Hernquist, L.\ 2005, \mnras, 361, 776
	\bibitem[Stern et al.(2012)]{2012ApJ...753...30S} Stern, D., Assef, R.~J., Benford, D.~J., et al.\ 2012, \apj, 753, 30
	\bibitem[Stoughton et al.(2002)]{2002AJ....123..485S} Stoughton, C., Lupton, R.~H., Bernardi, M., et al.\ 2002, \aj, 123, 485
	\bibitem[Tadhunter et al.(2005)]{2005MNRAS.356..480T} Tadhunter, C., Robinson, T.~G., Gonz{\'a}lez Delgado, R.~M., Wills, K., \& Morganti, R.\ 2005, \mnras, 356, 480 
	\bibitem[Taniguchi(1999)]{1999ApJ...524...65T} Taniguchi, Y.\ 1999, \apj, 524, 65
	\bibitem[Toomre \& Toomre(1972)]{1972ApJ...178..623T} Toomre, A., \& Toomre, J.\ 1972, \apj, 178, 623 
	\bibitem[Toomre(1977)]{1977ARA&A..15..437T} Toomre, A.\ 1977, \araa, 15, 437
	\bibitem[Tremaine et al.(2002)]{2002ApJ...574..740T} Tremaine, S., Gebhardt, K., Bender, R., et al.\ 2002, \apj, 574, 740
	\bibitem[Treister et al.(2012)]{2012ApJ...758L..39T} Treister, E., Schawinski, K., Urry, C.~M., \& Simmons, B.~D.\ 2012, \apjl, 758, L39 
	\bibitem[Torrey et al.(2012)]{2012ApJ...746..108T} Torrey, P., Cox, T.~J., Kewley, L., \& Hernquist, L.\ 2012, \apj, 746, 108
	\bibitem[Vanden Berk et al.(2001)]{2001AJ....122..549V} Vanden Berk, D.~E., Richards, G.~T., Bauer, A., et al.\ 2001, \aj, 122, 549 
	\bibitem[Vanden Berk et al.(2006)]{2006AJ....131...84V} Vanden Berk, D.~E., Shen, J., Yip, C.-W., et al.\ 2006, \aj, 131, 84
	\bibitem[Van Wassenhove et al.(2012)]{2012ApJ...748L...7V} Van Wassenhove, S., Volonteri, M., Mayer, L., et al.\ 2012, \apjl, 748, L7 
	\bibitem[Veilleux \& Osterbrock(1987)]{1987ApJS...63..295V} Veilleux, S., \& Osterbrock, D.~E.\ 1987, \apjs, 63, 295 
	\bibitem[Veilleux et al.(2002)]{2002ApJS..143..315V} Veilleux, S., Kim, D.-C., \& Sanders, D.~B.\ 2002, \apjs, 143, 315
	\bibitem[Voit(1992)]{1992MNRAS.258..841V} Voit, G.~M.\ 1992, \mnras, 258, 841 
	\bibitem[Walter et al.(2002)]{2002AJ....123..225W} Walter, F., Weiss, A., Martin, C., \& Scoville, N.\ 2002, \aj, 123, 225
	\bibitem[Wild et al.(2010)]{2010MNRAS.405..933W} Wild, V., Heckman, T., \& Charlot, S.\ 2010, \mnras, 405, 933
	\bibitem[Wei et al.(2013)]{2013ApJ...772...28W} Wei, P., Shang, Z., Brotherton, M.~S., et al.\ 2013, \apj, 772, 28 
	\bibitem[Wold et al.(2010)]{2010MNRAS.408..713W} Wold, I., Sheinis, A.~I., Wolf, M.~J., \& Hooper, E.~J.\ 2010, \mnras, 408, 713
	\bibitem[Wright et al.(2010)]{2010AJ....140.1868W} Wright, E.~L., Eisenhardt, P.~R.~M., Mainzer, A.~K., et al.\ 2010, \aj, 140, 1868-1881 
	\bibitem[Wu et al.(1998a)]{1998A&AS..127..521W} Wu, H., Zou, Z.~L., Xia, X.~Y., \& Deng, Z.~G.\ 1998, \aaps, 127, 521
	\bibitem[Wu et al.(1998b)]{1998A&AS..132..181W} Wu, H., Zou, Z.~L., Xia, X.~Y., \& Deng, Z.~G.\ 1998, \aaps, 132, 181
	\bibitem[Wu et al.(2007)]{2007ApJ...668...87W} Wu, H., Zhu, Y.-N., Cao, C., \& Qin, B.\ 2007, \apj, 668, 87
	\bibitem[Xia et al.(2002)]{2002ApJ...564..196X} Xia, X.~Y., Xue, S.~J., Mao, S., et al.\ 2002, \apj, 564, 196
	\bibitem[Yesuf et al.(2014)]{2014ApJ...792...84Y} Yesuf, H.~M., Faber, S.~M., Trump, J.~R., et al.\ 2014, \apj, 792, 84
	\bibitem[York et al.(2000)]{2000AJ....120.1579Y} York, D.~G., Adelman, J., Anderson, J.~E., Jr., et al.\ 2000, \aj, 120, 1579 
	\bibitem[Zibetti et al.(2013)]{2013MNRAS.428.1479Z} Zibetti, S., Gallazzi, A., Charlot, S., Pierini, D., \& Pasquali, A.\ 2013, \mnras, 428, 1479
	\bibitem[Zhang-hu \& Qiu-sheng(2016)]{2016ChA&A..40..291Z} Zhang-hu, C., \& Qiu-sheng, G.\ 2016, \caa, 40, 291
	\bibitem[Zheng et al.(1999)]{1999A&A...349..735Z} Zheng, Z., Wu, H., Mao, S., et al.\ 1999, \aap, 349, 735 
	\bibitem[Zou et al.(1991)]{1991MNRAS.252..593Z} Zou, Z., Xia, X., Deng, Z., \& Su, H.\ 1991, \mnras, 252, 593
\end{thebibliography}

\defcitealias{1998A&AS..127..521W}{Wu et al. 1998a}
\defcitealias{1998A&AS..132..181W}{Wu et al. 1998b}

\clearpage

\clearpage

\clearpage

\clearpage

\clearpage

\begin{figure}[h]
	\centering  
	\subfigure{
		\label{Fig.sub.1}
		\includegraphics[width=0.45\textwidth]{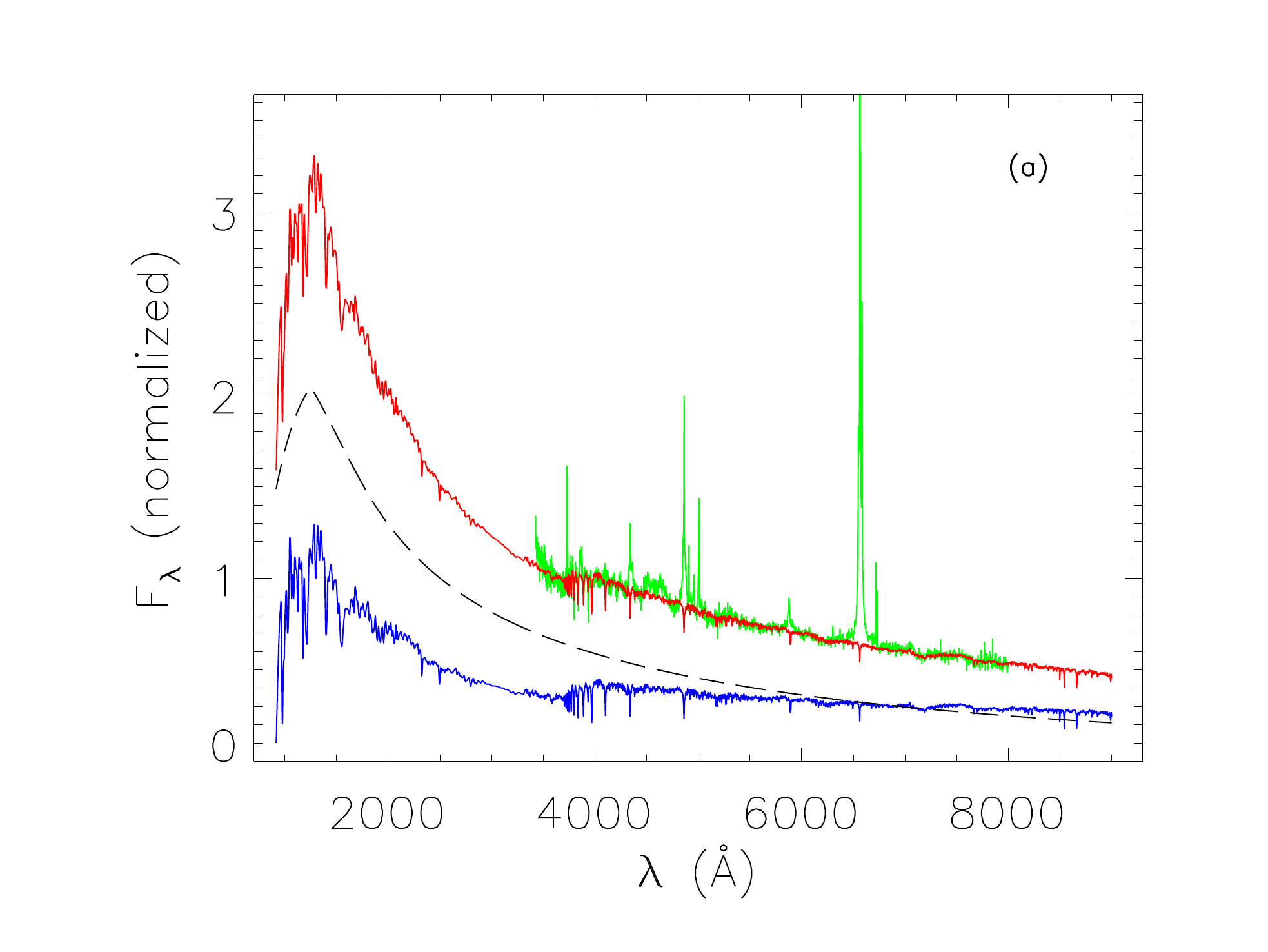}}
	\subfigure{
		\label{Fig.sub.2}
		\includegraphics[width=0.45\textwidth]{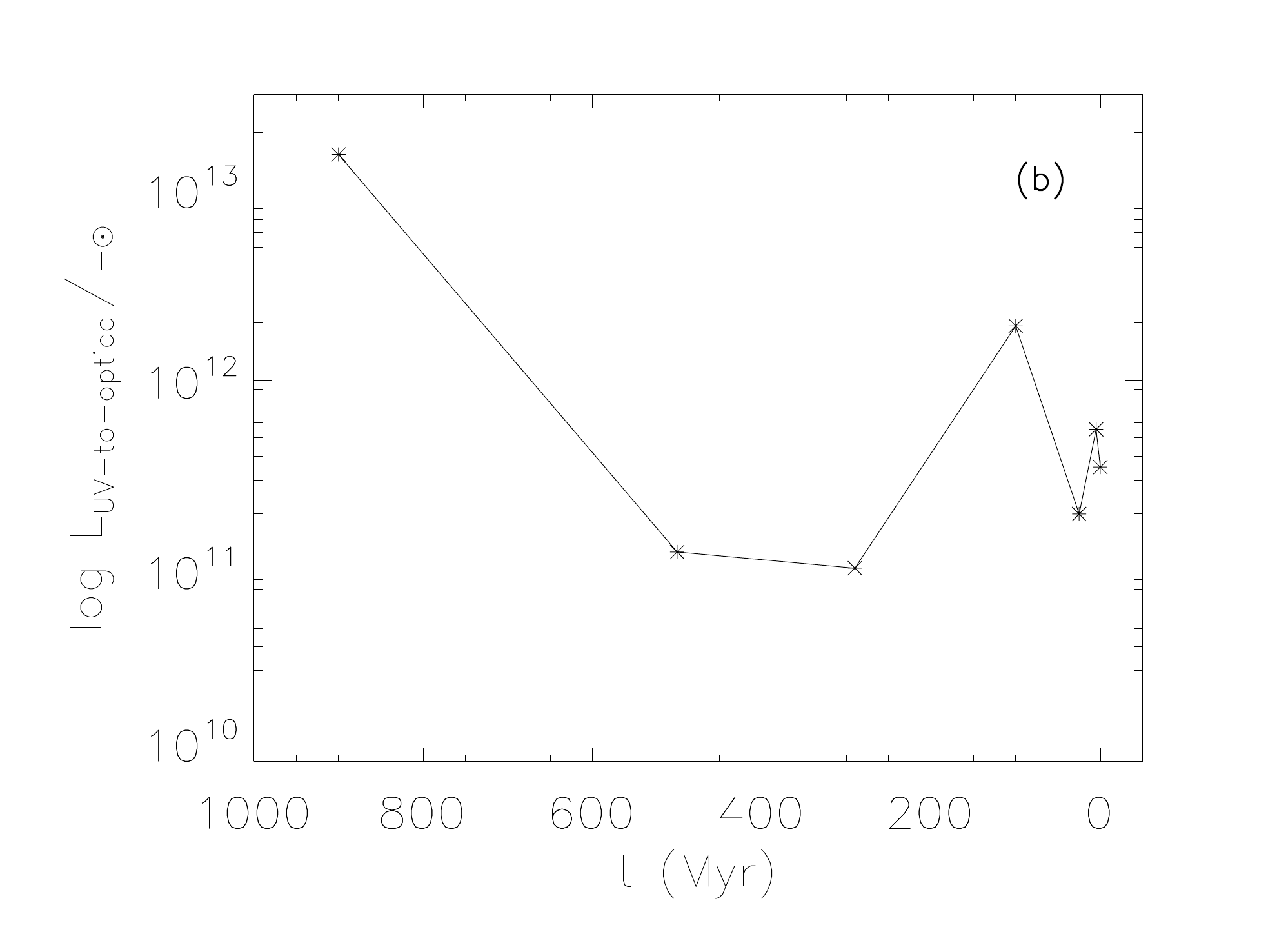}}
	\caption{Panel (a) shows the rebuilt model spectrum (red solid line) of SDSS J153705.95+005522.8,  which is superimposed by the observed one (green solid), the host galaxy spectrum (blue solid line) as well as the power-law spectrum (black dashed line). Panel (b) shows the SFH of the same host galaxy.}
	\label{Fig.main}
\end{figure}

\clearpage

\clearpage

\clearpage

\clearpage

\clearpage

\clearpage

\clearpage

\clearpage

\clearpage
	\centering
	\begin{longtable}[l]{llllllllc}
				\caption{The main parameter of of our source. (1) SDSS ID; (2-3) coordinates (4) redshift (5) WISE magnitude (6) Interaction}\\
		\hline 
		\hline 
SDSS\_name         & ra      & dec     & z      & W1     & W2     & W3    & W4    & Interaction \\
\hline 
101405.89+000620.3 & 10:14:06 & +00:06:22 & 0.140 & 11.639 & 10.573 & 7.956  & 5.899 &             \\
111807.47+002734.9 & 11:18:07 & +00:27:36 & 0.168 & 13.623 & 13.227 & 10.930 & 7.876 &             \\
113021.41+005823.0 & 11:30:21 & +00:58:23 & 0.132 & 12.514 & 11.696 & 8.829  & 6.232 &             \\
112852.59-032130.5 & 11:28:53 & -03:21:29 & 0.197 & 13.516 & 12.705 & 8.786  & 6.661 &             \\
125933.48-012833.3 & 12:59:34 & -01:28:34 & 0.266 & 13.154 & 12.260 & 9.108  & 6.595 & Y           \\
171411.63+575833.9 & 17:14:12 & +57:58:34 & 0.092 & 11.439 & 10.544 & 7.717  & 5.248 &             \\
025938.15+004216.3 & 02:59:38 & +00:42:18 & 0.195 & 13.678 & 13.128 & 9.856  & 7.348 &             \\
080037.62+461257.9 & 08:00:38 & +46:12:58 & 0.238 & 13.309 & 12.403 & 9.127  & 6.478 & Y           \\
090906.40+535040.4 & 09:09:06 & +53:50:42 & 0.273 & 13.826 & 13.115 & 9.956  & 7.826 &             \\
034831.88-071145.9 & 03:48:32 & -07:11:46 & 0.183 & 12.591 & 11.640 & 8.557  & 6.055 & N           \\
090158.88+002313.8 & 09:01:59 & +00:23:13 & 0.196 & 12.818 & 11.885 & 8.743  & 6.213 & Y           \\
111713.91+674122.7 & 11:17:14 & +67:41:24 & 0.247 & 12.697 & 11.695 & 8.721  & 6.323 & N           \\
131953.15+033335.9 & 13:19:53 & +03:33:36 & 0.208 & 13.577 & 12.995 & 9.865  & 7.857 &             \\
133715.92+030936.5 & 13:37:16 & +03:09:36 & 0.192 & 13.328 & 12.574 & 9.314  & 7.085 &             \\
150420.90+015159.3 & 15:04:21 & +01:51:58 & 0.182 & 12.958 & 11.960 & 8.644  & 6.068 &             \\
075057.26+353037.6 & 07:50:57 & +35:30:36 & 0.175 & 13.355 & 12.172 & 8.639  & 6.327 &             \\
082405.19+445246.0 & 08:24:05 & +44:52:44 & 0.219 & 13.397 & 12.456 & 8.715  & 6.604 &             \\
104451.87+035251.9 & 10:44:52 & +03:52:52 & 0.206 & 13.291 & 12.482 & 9.372  & 7.209 &             \\
151600.39+572415.7 & 15:16:00 & +57:24:14 & 0.204 & 12.948 & 12.108 & 9.255  & 6.878 &             \\
162633.92+480230.1 & 16:26:34 & +48:02:31 & 0.242 & 12.904 & 11.977 & 8.869  & 6.666 & N           \\
003657.17-100810.6 & 00:36:57 & -10:08:10 & 0.187 & 13.208 & 12.554 & 9.164  & 6.627 & N           \\
033156.88+002605.2 & 03:31:57 & +00:26:06 & 0.236 & 13.602 & 12.761 & 9.982  & 7.696 &             \\
113630.11+621902.4 & 11:36:30 & +62:19:01 & 0.211 & 13.511 & 12.715 & 9.388  & 6.832 & Y           \\
083453.39+384708.5 & 08:34:53 & +38:47:10 & 0.184 & 13.445 & 12.611 & 9.759  & 7.746 &             \\
083917.34+392817.9 & 08:39:17 & +39:28:19 & 0.186 & 13.182 & 12.376 & 9.540  & 7.285 &             \\
133706.93+051803.3 & 13:37:07 & +05:18:04 & 0.163 & 13.820 & 13.342 & 9.643  & 6.762 & Y           \\
081116.70+320935.3 & 08:11:17 & +32:09:36 & 0.153 & 12.544 & 11.743 & 9.215  & 6.648 & Y           \\
110051.02+513502.1 & 11:00:51 & +51:35:02 & 0.213 & 13.254 & 12.452 & 9.547  & 7.350 & N           \\
123915.40+531414.6 & 12:39:15 & +53:14:13 & 0.201 & 12.999 & 12.407 & 9.637  & 7.070 & Y           \\
081835.59+390911.1 & 08:18:36 & +39:09:11 & 0.186 & 13.623 & 12.804 & 10.140 & 7.488 & N           \\
132832.58-023321.4 & 13:28:33 & -02:33:22 & 0.183 & 12.955 & 12.349 & 9.764  & 7.200 & N           \\
134452.60-011452.2 & 13:44:53 & -01:14:53 & 0.177 & 13.210 & 12.443 & 9.449  & 6.786 &             \\
081438.27+290619.9 & 08:14:38 & +29:06:22 & 0.225 & 13.565 & 12.646 & 9.609  & 7.017 &             \\
105705.40+580437.4 & 10:57:06 & +58:04:37 & 0.140 & 12.514 & 11.889 & 8.863  & 6.736 & Y           \\
131750.32+601040.8 & 13:17:50 & +60:10:41 & 0.136 & 12.180 & 11.324 & 8.403  & 5.799 & Y           \\
133237.93+593053.6 & 13:32:38 & +59:30:54 & 0.171 & 12.234 & 11.466 & 8.495  & 5.927 & Y           \\
133435.38+575015.6 & 13:34:35 & +57:50:17 & 0.123 & 12.309 & 11.411 & 8.442  & 6.038 &             \\
171756.03+261148.6 & 17:17:56 & +26:11:49 & 0.145 & 13.791 & 13.022 & 10.180 & 7.919 & N           \\
152008.23+461615.3 & 15:20:08 & +46:16:16 & 0.176 & 13.491 & 12.748 & 10.180 & 8.176 & Y           \\
134615.88+580008.1 & 13:46:16 & +58:00:07 & 0.162 & 12.936 & 12.199 & 9.381  & 6.576 &             \\
151907.33+520605.9 & 15:19:07 & +52:06:07 & 0.137 & 11.042 & 9.848  & 6.528  & 3.947 & Y           \\
154518.05+463837.9 & 15:45:18 & +46:38:38 & 0.228 & 11.385 & 10.328 & 7.395  & 4.752 & Y           \\
093302.68+385228.0 & 09:33:03 & +38:52:26 & 0.177 & 12.450 & 11.618 & 8.532  & 6.363 & Y           \\
100302.15+095832.8 & 10:03:02 & +09:58:34 & 0.253 & 13.420 & 12.416 & 9.687  & 7.535 & N           \\
125908.35+561530.7 & 12:59:08 & +56:15:32 & 0.160 & 12.371 & 11.444 & 8.791  & 6.214 &             \\
143123.52+392501.4 & 14:31:24 & +39:25:01 & 0.161 & 13.349 & 12.718 & 9.722  & 7.479 &             \\
113651.66+445016.4 & 11:36:52 & +44:50:17 & 0.115 & 12.302 & 11.630 & 8.373  & 6.254 &             \\
135719.47+394045.3 & 13:57:19 & +39:40:44 & 0.265 & 13.566 & 12.767 & 9.361  & 7.381 &             \\
140007.29+405357.6 & 14:00:07 & +40:53:56 & 0.167 & 12.891 & 12.117 & 9.471  & 6.807 & Y           \\
105409.18+412827.6 & 10:54:09 & +41:28:26 & 0.230 & 14.781 & 14.057 & 11.210 & 8.469 &             \\
112930.76+431017.3 & 11:29:31 & +43:10:16 & 0.186 & 13.282 & 12.729 & 9.991  & 7.636 & N           \\
091020.11+312417.8 & 09:10:20 & +31:24:18 & 0.265 & 13.666 & 12.751 & 9.480  & 7.043 & Y           \\
114926.47+112629.0 & 11:49:26 & +11:26:28 & 0.177 & 12.941 & 12.119 & 9.490  & 7.690 & N           \\
121945.03+082117.9 & 12:19:45 & +08:21:18 & 0.228 & 13.144 & 12.133 & 8.772  & 6.095 & Y           \\
155654.47+253233.6 & 15:56:54 & +25:32:35 & 0.164 & 13.253 & 12.604 & 10.090 & 7.629 & Y           \\
155958.01+261102.7 & 15:59:58 & +26:11:02 & 0.228 & 13.374 & 12.465 & 9.411  & 6.869 & Y           \\
160700.93+245056.6 & 16:07:01 & +24:50:56 & 0.183 & 12.720 & 11.707 & 8.529  & 5.977 & Y           \\
132105.98+504634.4 & 13:21:06 & +50:46:34 & 0.233 & 12.881 & 11.945 & 8.588  & 6.320 &             \\
141557.25+495334.5 & 14:15:57 & +49:53:35 & 0.185 & 13.558 & 12.781 & 9.119  & 6.612 & Y           \\
134704.91+144137.6 & 13:47:05 & +14:41:38 & 0.134 & 12.509 & 11.614 & 8.370  & 5.828 &             \\
151453.27+053636.8 & 15:14:53 & +05:36:36 & 0.173 & 12.844 & 11.934 & 9.030  & 6.464 &             \\
080652.11+564412.7 & 08:06:52 & +56:44:13 & 0.180 & 12.027 & 10.848 & 7.822  & 5.476 & Y           \\
100208.14+345353.7 & 10:02:08 & +34:53:53 & 0.205 & 12.718 & 11.642 & 8.736  & 6.284 & N           \\
122028.07+405035.0 & 12:20:28 & +40:50:35 & 0.221 & 12.955 & 12.129 & 9.306  & 7.021 & Y           \\
130712.33+340622.5 & 13:07:12 & +34:06:22 & 0.147 & 12.686 & 11.607 & 8.205  & 5.422 &             \\
115515.86+380234.9 & 11:55:16 & +38:02:35 & 0.143 & 12.825 & 11.983 & 8.806  & 6.404 & Y           \\
115828.53+373450.1 & 11:58:29 & +37:34:52 & 0.186 & 13.824 & 13.041 & 9.416  & 7.038 & Y           \\
121006.01+333602.9 & 12:10:06 & +33:36:04 & 0.225 & 13.364 & 12.513 & 9.810  & 7.512 &             \\
135852.46+295413.1 & 13:58:53 & +29:54:14 & 0.113 & 11.426 & 10.446 & 7.591  & 5.128 &             \\
142230.34+295224.2 & 14:22:30 & +29:52:23 & 0.113 & 12.789 & 12.005 & 7.529  & 4.402 & Y           \\
151337.07+201133.6 & 15:13:37 & +20:11:35 & 0.270 & 12.932 & 11.961 & 8.959  & 6.438 & N           \\
161002.70+202108.5 & 16:10:03 & +20:21:07 & 0.217 & 13.166 & 12.419 & 9.414  & 7.149 & N           \\
110805.03+271313.9 & 11:08:05 & +27:13:16 & 0.358 & 13.148 & 12.221 & 9.325  & 7.133 &             \\
091848.61+211717.0 & 09:18:49 & +21:17:17 & 0.149 & 11.063 & 10.039 & 7.320  & 4.824 & Y           \\
102955.58+244523.2 & 10:29:56 & +24:45:22 & 0.220 & 13.789 & 13.018 & 10.340 & 7.848 & Y           \\
115248.18+212255.5 & 11:52:48 & +21:22:55 & 0.171 & 13.108 & 12.568 & 9.734  & 7.408 & Y           \\
154526.04+141159.3 & 15:45:26 & +14:12:00 & 0.284 & 13.552 & 12.545 & 9.226  & 6.692 &             \\
132954.86+182041.7 & 13:29:55 & +18:20:42 & 0.188 & 13.266 & 12.490 & 9.253  & 6.827 &             \\
150408.46+143123.3 & 15:04:08 & +14:31:23 & 0.118 & 11.667 & 10.657 & 7.207  & 4.914 & Y           \\
153031.25+120734.0 & 15:30:31 & +12:07:34 & 0.197 & 13.517 & 12.800 & 10.260 & 8.115 &             \\
152205.06+012626.6 & 15:22:05 & +01:26:28 & 0.113 & 11.871 & 10.962 & 7.493  & 5.045 &             \\
153705.95+005522.8 & 15:37:06 & +00:55:23 & 0.136 & 12.497 & 11.674 & 8.314  & 6.232 &             \\ \hline 
\label{table:observation}
	\end{longtable}

\clearpage
\begin{table}[]
	\caption{The KS-test of WISE color.     } 	
	
	\begin{tabular}{|c|cc|}
		\hline 
		type   & W12 & W23 \\
		\hline 
		Working sample vs Control  & $\ll 0.001$    & $\ll 0.001$    \\
		typeO+typeB     vs    typeA+typeAB  &        & $\ll 0.001$     \\
		\hline 
	\end{tabular}
	\label{table:KS_WISE}
\end{table}

\clearpage
\begin{table}[h]
	\centering
\caption{The mean values with standard errors of BH mass and Eddington ratio for four types and control sample. }

\begin{tabular}{|c|cc|cc|}
		\hline 
				   & &logBH   & &logEdd \\
	\hline 	
		type   & mean & standard errors & mean & standard errors \\
		\hline 
		typeO  & 8.65    & 0.12  & -1.63    & 0.12   \\
		typeA  & 8.45      & 0.10  & -1.41      & 0.09   \\
		typeB  & 8.67       & 0.13  & -1.60      & 0.15   \\
		typeAB & 8.20   & 0.09   & -1.09    & 0.10  \\
		control sample &8.23& 0.02   &  -0.56 &   0.02\\

	\hline
	\end{tabular}
	\label{table:mean_SE}
\end{table}

\end{document}